\documentclass[useAMS,usenatbib]{mnras}

\usepackage{graphicx}	
\usepackage{subcaption}
\usepackage{floatrow}
\usepackage{amsmath}	
\usepackage{amssymb}	
\usepackage{multicol}        
\usepackage{bm}		

\newcommand{\vect}[1]{\boldsymbol{#1}}

\usepackage{color}
\definecolor{orange}{rgb}{0.90,0.60,0}
\definecolor{skyblue}{rgb}{0.35,0.70,0.90}
\definecolor{green}{rgb}{0,0.60,0.50}
\definecolor{yellow}{rgb}{0.95,0.90,0.25}
\definecolor{blue}{rgb}{0,0.45,0.70}
\definecolor{vermilion}{rgb}{0.80,0.40,0}
\definecolor{lilac}{rgb}{0.80,0.60,0.70}

\newcommand{\dif}{\mathrm{d}}

\newcommand{\yr}{\,\mathrm{yr}}
\newcommand{\Myr}{\,\mathrm{Myr}}
\newcommand{\Gyr}{\,\mathrm{Gyr}}
\newcommand{\AU}{\,\mathrm{au}}
\newcommand{\pc}{\,\mathrm{pc}}
\newcommand{\ME}{M_{\oplus}}

\newcommand{\MJ}{M_{\rm J}}

\newcommand{\MSol}{M_{\odot}}

\newcommand{\nhat}{\vect{\hat{n}}}

\newcommand{\khat}{\vect{\hat{k}}}

\defcitealias{HT1986}{HT86}

\title[White-dwarf pollution from exocomets]{On the pollution of white dwarfs by exo-Oort cloud comets}

\author[O'Connor, Lai \& Seligman]
{Christopher E.\ O'Connor$^{1,2}$\thanks{E-mail: coconnor@astro.cornell.edu}, Dong Lai$^{1,3}$, and Darryl Z.\ Seligman$^{1,3}$ \\
$^{1}$Department of Astronomy and Cornell Center for Astrophysics and Planetary Science, Cornell University, Ithaca, NY 14853, U.S.A. \\
$^{2}$Kavli Institute for Theoretical Physics, University of California, Santa Barbara, CA 93106, U.S.A. \\
$^{3}$Carl Sagan Institute, Cornell University, Ithaca, NY 14853, U.S.A.
}

\begin{document}

\date{Submitted 9 June 2023.}

\pagerange{\pageref{firstpage}--\pageref{lastpage}} \pubyear{2023}

\maketitle

\label{firstpage}

\begin{abstract}
    A large fraction of white dwarfs (WDs) have metal-polluted atmospheres, 
    which are produced by accreting material from remnant planetary systems. 
    The composition of the accreted debris broadly resembles that of rocky Solar System objects. 
    Volatile-enriched debris with compositions similar to long-period comets (LPCs) is rarely observed. 
    We attempt to reconcile this dearth of volatiles with the premise 
    that exo-Oort clouds (XOCs) occur around a large fraction of planet-hosting stars. 
    We estimate the comet accretion rate from an XOC analytically, adapting the `loss cone' theory of LPC delivery in the Solar System. 
    We investigate the dynamical evolution of an XOC during late stellar evolution. 
    Using numerical simulations, we show that 1 to 30 per cent of XOC objects remain bound 
    after anisotropic stellar mass loss imparting a WD natal kick of $\sim 1 \, {\rm km \, s^{-1}}$. 
    We also characterize the surviving comets’ distribution function. 
    Surviving planets orbiting a WD can prevent the accretion of XOC comets by the star. 
    A planet's `dynamical barrier' is effective at preventing comet accretion 
    if the energy kick imparted by the planet exceeds the comet's orbital binding energy. 
    By modifying the loss cone theory, we calculate the amount by which a planet reduces the WD’s accretion rate. 
    We suggest that the scarcity of volatile-enriched debris in polluted WDs 
    is caused by an unseen population of $10$--$100 \AU$ scale giant planets acting as barriers to incoming LPCs. 
    Finally, we constrain the amount of volatiles delivered to a planet in the habitable zone of an old, cool WD.
\end{abstract}


\section{Introduction}

The Solar System has two major populations of comets. 
Long-period comets (period $P \geq 200 \yr$; LPCs) have nearly parabolic orbits and isotropic inclinations. 
They originate from the Oort cloud (OC), a swarm of $\sim 10^{11}$--$10^{12}$ bodies 
\citep{Oort1950, Francis2005, BM2013, Boe+2019} 
with a combined mass of $\approx 1$--$20 \ME$ 
\citep{Weissman1996, FB2000, Francis2005, Brasser2008, Boe+2019} 
extending up to $\sim 10^{5} \AU$ from the Sun.  
LPCs most likely have a broad distribution of eccentricities, 
but only those with eccentricity $\simeq 1$ are observed entering the inner Solar System. 
Short-period comets ($P < 200 \yr$; SPCs) typically have low inclinations relative to the ecliptic plane. 
Many come from the trans-Neptunian region \citep[e.g.][]{DL1997, Levison+2006, VM2008}, 
but the Halley-type SPCs may be former OC members captured at shorter periods \citep{Levison+2001, Nesvorny+2017}. 

The small bodies within the Solar System -- including the comets -- are believed to be 
byproducts of the system's formation and subsequent dynamical evolution 
\citep[see the recent review by][]{KV2022}. 
Numerous icy planetesimals formed in or beyond the vicinity of the giant planets (when the latter formed, although they have since migrated) 
and were subsequently expelled by gravitational scattering 
\citep[e.g.][]{Duncan+1987, HM1999, BM2013, VND2019}. Most were ejected into interstellar space, 
but 1 to 10 per cent were captured on relatively stable orbits in the OC
by external perturbations, such as stellar flybys and the Galactic tide 
\citep{HM1999, Dones+2004, Brasser+2010, HK2015}. 
Planetesimals expelled from the giant-planet region also populated the Kuiper belt's scattered disc, 
the main source of SPCs. 
The existence of extrasolar comet reservoirs is therefore a natural extrapolation 
from the ubiquity of extrasolar planetary systems with long-period giant planets 
\citep{Suzuki+2016, Fernandes+2019, Poleski+2021, Fulton+2021}. 
In this work, we focus on extrasolar Oort clouds (XOCs) specifically.

Two lines of evidence suggest exocomets may be common. 
One comes from rapidly varying spectroscopic absorption features in stellar observations, 
which are naturally explained by evaporating bodies falling onto their host stars. 
This occurs most famously in the $\beta$~Pic system \citep{Ferlet+1987, Beust+1991, Pavlenko+2022}. 
The other line of evidence comes from photometric transits of dusty cometary debris
\citep{Boyajian+2016, Kiefer+2017, Rappaport+2018, Kennedy+2019, Zieba+2019}. 
The extent to which the observed exocomets are analogous to Solar System comets is unclear 
(see \citealt{Strom+2020} for an introductory review of this topic).
The size-frequency distribution of exocomets transiting $\beta$~Pic 
is similar to those in the Solar System population \citep{Etangs2022}. 
However, relatively little is known about exocometary chemical compositions 
\citep[e.g.][]{ZS2012, Matra+2015, Kral+2017}. 

Interstellar interlopers such as 1I/`Oumuamua and 2I/Borisov  
provide an indirect hint at the existence of XOCs 
because they may represent the large number of ejected planetesimals implied by XOC formation models
\citep{Gaidos+2017, Gaidos2018, Do+2018, MoroMartin2018, MoroMartin2019, PZ+2021a, Seligman+2022b, JewittSeligman2022}. 
They also add to the available compositional information about exocomets. 
For example, 2I/Borisov was enriched in CO relative to H$_2$O, 
indicative of formation at the CO snowline or beyond \citep{Bodewits2020,Cordiner2020,yang2021}. 
However, absent spectroscopic measurements of interstellar comets in the inner Solar System, 
measuring compositional abundances of XOCs is difficult.

Observations of white dwarfs (WDs) provide a promising alternative method 
to identify XOCs and measure their composition. 
Between 25 and 50 per cent of WDs exhibit trace amounts of externally derived metals in their atmospheres \citep{Zuckerman+2003, Zuckerman+2010, Barstow+2014, KGF2014, Wilson+2019}. 
This subset is commonly referred to as polluted WDs. 
The pollution is generally attributed to continual accretion of debris from a dynamically evolving planetary system 
\citep[e.g.][]{DS2002, DWS2012, FH2014, PM2017, SNZ2017, Mustill+2018, Maldonado+2020, Li+2022, OTL2022, Trierweiler+2022}. 
The cumulative mass and chemical composition of the accreted material 
provide information about the parent body or bodies 
\citep[e.g.][]{Zuckerman+2007, Xu+2019, DDY2021, Buchan+2022, Trierweiler+2023}. 
Polluted WDs are therefore important probes of extrasolar planetesimal populations.

The characterization of parent bodies producing WD pollution is subtle and subject to uncertainties 
related to the mixing and settling of metals within the WD's atmosphere. 
The simplest and most widely implemented methodology accounts only
for gravitational settling  \citep{Dupuis+1993}. 
The settling time-scale varies significantly along the WD cooling sequence, 
taking values of $\sim 10^{-2}$--$10^{3} \yr$ for a hydrogen-rich atmosphere 
and $\sim 10^{4}$--$10^{6} \yr$ for a helium-rich atmosphere 
\citep[e.g.][]{Koester2009, BB2019}. 
Accounting for additional processes such as thermohaline mixing \citep{Deal+2013, Wachlin+2017, BB2018} 
and convective overshooting \citep{Tremblay+2015, Tremblay+2017, BB2019, Cunningham+2019} 
can alter the total metal accretion rate inferred for a given WD. 
However, these effects do not alter the inferred composition of the accreted material. 
Radiative levitation can also affect a WD's metal abundances and inferred accretion rate \citep[e.g.][]{Chayer+1995, Chayer+1995b, Chayer2014}, 
but only for effective temperatures above $20\,000 \, {\rm K}$. 

To date, a few dozen polluted WDs have been the subject of spectroscopic follow-up studies measuring their elemental abundances \citep{JY2014, ZY2018}. 
In general, the accreted material is dominated by the silicate-forming elements O, Mg, Si, and Fe. 
The measured abundance ratios broadly resemble those found in the CI chondrites 
(the most pristine meteorites in terms of composition)
and the bulk Earth \citep[e.g.][]{Zuckerman+2007, Xu+2019, Doyle+2023, Trierweiler+2023}, 
suggesting that rocky parent bodies are the predominant source of pollution. 
Debris from an icy parent body such as a comet would also be rich in C and N, 
producing higher stellar abundances of these volatile elements than are generally observed. 
For example, Halley's comet is enriched in C and N by $\gtrsim 1 \, {\rm dex}$ by number 
relative to the CI chondrites \citep{Jessberger+1988, Lodders2021}. 
At present, the only known polluted WD with C and N abundances 
comparable to Halley's comet is WD\,1425+540 \citep{Xu+2017}. 
This WD is a member of a wide binary system, 
so gravitational perturbations from the companion star
may facilitate the delivery of volatile-enriched icy objects \citep[e.g.][]{BV2015, SNZ2017}. 
Some polluted WDs appear to have accreted water-rich bodies, 
indicated by the detection of trace hydrogen in helium-dominated atmospheres
\citep[e.g.][]{GF+2017, HGK2018, Coutu+2019, Hoskin+2020, Izquierdo+2021}. 
\citet{Veras+2014} proposed that this can be explained by occasional comet impacts. 
On the whole, chemical evidence indicates that volatile-enriched exocomets 
are not the main source of pollution in most WD systems.

In this paper, we address the question of why accretion of volatile-enriched debris is rare among WDs. 
\citet*{AFS1986} considered this question and argued that the overall fraction of single stars with XOCs is small. 
However, their conclusion was based on a limited sample of polluted WDs discovered at the time. 
Moreover, it is now in tension with the apparent ubiquity of volatile-enriched interstellar comets. 
An alternative explanation is that the dynamical evolution of XOCs around planet-hosting WDs 
reduces the rate of comet accretion events such that the star's volatile abundances remain below the minimum observable level. 
For WDs cooler than $20\,000 \, {\rm K}$, the minimum observable accretion rate for rocky debris is a few $10^{5} \, {\rm g \, s^{-1}}$ \citep{KGF2014, BX2022}. 
We assume that the same detection threshold applies to debris from volatile-enriched exocomets. 

In this work, we investigate dynamical processes that reduce the rate 
of comet accretion events for single WDs hosting XOCs. 
In Section \ref{s:OC-Dynamics}, we apply the standard loss cone theory of the delivery of LPCs
to estimate the comet accretion rate from a Solar-System-like XOC. 
In Section \ref{s:MassLoss}, we use numerical simulations 
to study the retention of comets during late stellar evolution. 
In Section \ref{s:Planets}, we consider how a surviving planetary system around a WD 
modifies the rate of comet bombardment. 
In Section \ref{s:Discussion}, we synthesize the results of the preceding sections, 
compare our results with those of related works, 
and provide some qualifications of our conclusions. 
We also discuss the potential relevance of comets
to the question of planetary habitability around WDs. 
We summarize our main results in Section \ref{s:Conclusion}.

\section{Dynamics of comet injection} \label{s:OC-Dynamics}

The dynamical evolution of OC comets is governed by a combination of several effects. 
These include secular torques due to both the Galactic tidal (GT) field and an inner planetary system 
(\citealt{HT1986}; Section \ref{s:Planets:Secular}) 
and random, impulsive perturbations from stellar flybys \citep[e.g.][]{Oort1950, Hills1981, HTA1987}. 
External perturbations have contributed to the gradual erosion 
of the OC in the Solar System over its $4.5 \Gyr$ lifetime. 
Comets with low perihelion distances can be scattered during encounters with planets \citep[e.g.][]{Fernandez1981}. 
These processes lead to either the ejection of comets as free-floating bodies 
or their capture in the inner Solar System as SPCs. 
Planets can also tidally disrupt or accrete passing comets, 
as in Comet Shoemaker--Levy 9's collision with Jupiter \citep{Chapman1993}. 
Comets passing close to the Sun `fade' over successive passages \citep{Whipple1962, WT1999, BW2015} 
or are tidally disrupted. 

In the Solar System, the structure of the OC has presumably reached a quasi-steady state 
after a few Gyr of dynamical relaxation \citep[e.g.][]{Duncan+1987,HK2015}. 
The progenitors of polluted WDs are generally more massive than the Sun 
($\approx 1.5$--$3 \MSol$; \citealt{KGF2014}) 
and therefore have shorter main-sequence lifetimes. 
A cometary reservoir surrounding a star whose main-sequence lifetime 
is shorter than the reservoir's relaxation time 
may differ in structure from the OC. 
Differences between the architectures of the host's planetary system and the Solar System, 
as well as the galactic environments in which the systems formed, 
can also produce differently structured XOCs \citep{Fernandez1997}.
Nevertheless, the inferred properties of the OC provide 
a useful starting point for studying XOCs. 

In the following, we use the loss cone theory to determine the injection rate of comets from an XOC \citep{HT1986}. 
First, we introduce the quantities describing the orbits of comets and their distribution within the OC. 
The mass of the central star is denoted $M$. 
We assume that each comet is a mass-less test particle.
We denote the usual Keplerian elements as follows: 
semi-major axis $a$, eccentricity $e$, inclination $I$, argument of periastron $\chi$, 
longitude of the ascending node $\Omega$, and mean anomaly $l$. 
Related quantities are the periastron and apoastron distances $q = a(1-e)$ and $Q = a(1+e)$, 
mean motion $\omega = (G M / a^{3})^{1/2}$, and orbital period $P =  2 \pi / \omega$.
We also refer to the following Delaunay action variables:
\begin{equation} \label{eq:def_LJJz}
    L = (G M a)^{1/2}, \hspace{0.25cm}
    J = L (1 - e^{2})^{1/2}, \hspace{0.25cm}
    J_{z} = J \cos{I}.
\end{equation}
These are canonically conjugate to $l$, $\chi$, and $\Omega$ respectively.
The action $L$ is related to the orbital energy per unit mass, 
whilst $J$ and $J_{z}$ are the magnitude and vertical component of 
the orbital angular momentum per unit mass.
The distribution function $f$ of comets in an XOC specifies 
the number of comets per volume element of phase space, i.e.
\begin{equation}
    \dif N = f(L, J, J_{z}, l, \chi, \Omega) \, \dif L \, \dif J \, \dif J_{z} \, \dif l \, \dif \chi \, \dif \Omega.
\end{equation}
In principle, $f$ may also depend explicitly on the time $t$. 
We neglect this possibility for simplicity; 
this is equivalent to assuming the OC is dynamically relaxed.

\subsection{Galactic tides} \label{s:OC-Dynamics:Gtide}

\citet[][hereafter HT86]{HT1986} presented an analytic model of the secular evolution 
of a star--comet system perturbed by the tidal field of the Galaxy. 
In their model, the primary star is assumed to follow a circular orbit in the Galactic midplane. 
The Galactic potential $\Phi_{\rm GT}$ near the star can be approximated 
as the potential inside a slab with uniform density $\rho_{\rm g}$, i.e.
\begin{equation}
    \Phi_{\rm GT} = 2 \pi G \rho_{\rm g} z^{2},
\end{equation}
where $z$ is the vertical distance from the Galactic midplane.
We adopt a fiducial value $\rho_{\rm g} = 0.1 \MSol \pc^{-3}$ throughout this work \citep[e.g.][]{McKee+2015}. 
\citetalias{HT1986} considered the secular evolution of the comet, 
obtained by averaging $\Phi_{\rm GT}$ over the Keplerian orbit before applying Hamilton's equations. 
For the purposes of this paper, we require only the equations of motion for $J$ and $\chi$:
\begin{subequations} \label{eq:GT_EqsMot}
\begin{align}
    \frac{\dif J}{\dif t} &= - \frac{5}{2} \omega_{\rm GT} L \left( 1 - \frac{J^{2}}{L^{2}} \right) \left( 1 - \frac{J_{z}^{2}}{J^{2}} \right) \sin{(2 \chi)}, \label{eq:dJdt} \\
    \frac{\dif \chi}{\dif t} &= \omega_{\rm GT} \frac{J}{L} \left[ 1 - 5 \left( 1 - \frac{J_{z}^{2} L^{2}}{J^{4}} \right) \sin^{2}{\chi} \right]\,. \label{eq:dchidt}
\end{align}
\end{subequations}
In Eq.\ (\ref{eq:GT_EqsMot}), the characteristic oscillation frequency $\omega_{\rm GT}$ is given by,
\begin{align}
    \omega_{\rm GT} &= G \rho_{\rm g} P \\
    &\approx \frac{1}{2.2 \Gyr} \left( \frac{\rho_{\rm g}}{0.1 \MSol \pc^{-3}} \right) \left( \frac{a}{10^{4} \AU} \right)^{3/2} \left( \frac{M}{\MSol} \right)^{-1/2}. \nonumber
\end{align}

\subsubsection{Comet injection rate at small distances}

\citetalias{HT1986} calculated the rate at which comets are injected by the Galactic tide 
within a critical distance $q_{\rm cr} \ll a$ of the central star 
from a spherically symmetric cloud with distribution function $f(L,J)$. 
Their purpose was to estimate the rate at which OC comets 
are captured within the inner Solar System via encounters with the planets. 
In the absence of gravitational perturbations from a planetary system, 
we may use their result to estimate the flux of OC comets at arbitrarily small distances. 

It is convenient to denote the orbital angular momentum of an object 
with perihelion distance $q_{\rm cr}$ as 
\begin{equation}
    J_{\rm cr} \equiv (2 G M q_{\rm cr})^{1/2}
\end{equation}
with $J_{\rm cr} \ll L$. 
For a given $J_{\rm cr}$, we denote the injection rate of comets between $L$ and $L + \dif L$ as $\Gamma(L) \, \dif L$ 
and the total rate as
\begin{equation} \label{eq:Gamma_tot}
    \Gamma_{\rm tot} = \int_{0}^{\infty} \Gamma(L) \, \dif L.
\end{equation}
The phase-space region $J \leq J_{\rm cr}$ is called the ``loss cone'' (or ``loss cylinder'') 
because comets passing through this region are assumed to be removed from the XOC. 

\begin{figure}
    \centering
    \includegraphics[width=\columnwidth]{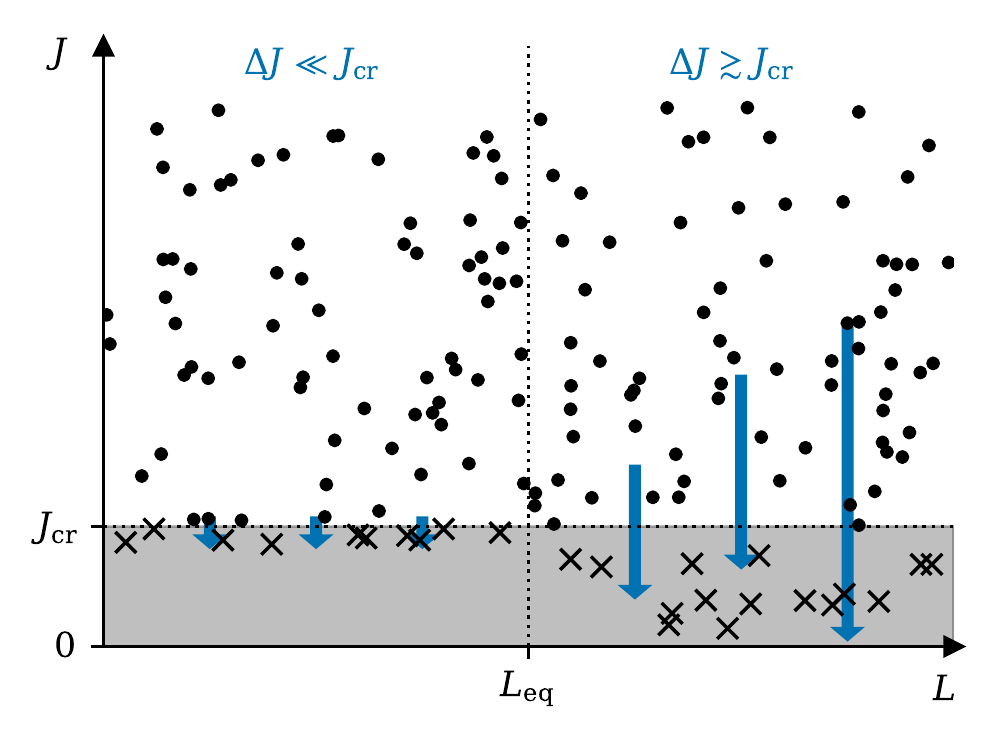}
    \caption{Schematic depiction of the loss cone theory of comet injection. 
    Blue arrows represent the typical change of a comet's angular momentum per orbit ($\Delta J$). 
    For $\Delta J \ll J_{\rm cr}$ ($L \lesssim L_{\rm eq}$), XOC comets (dark points) are injected 
    by diffusing slowly across the boundary of the cone ($J = J_{\rm cr}$); 
    the interior of the loss cone (shaded region) is empty 
    because the comets are lost within one orbital period (crosses). 
    For $\Delta J \gtrsim J_{\rm cr}$ ($L \gtrsim L_{\rm eq}$), 
    comets leap over the boundary and fill the loss cone before being lost.}
    \label{fig:losscone_basic}
\end{figure}

There are two regimes of the comet injection process, 
called the ``empty loss cone'' and ``filled loss cone.'' 
These correspond to different expressions for $\Gamma(L)$ in particular ranges of $L$. 
The loss cone is said to be empty when $\Delta J \lesssim J_{\rm cr}$, 
where $\Delta J \sim |\dot{J}| P$ is the typical change of $J$ per orbit. 
The injection rate for an empty loss cone is \citepalias{HT1986}
\begin{equation} \label{eq:dif_Gamma_empty}
    \Gamma_{\rm e}(L) \, \dif L \simeq \frac{160 \pi^{3} \rho_{\rm g} L^{4} J_{\rm cr}}{3 G M^{2}} f(L,J_{\rm cr}) \, \dif L,
\end{equation}
where $J_{\rm cr} \ll L$. 
This is calculated as the rate at which comets with $J \in [J_{\rm cr}, J_{\rm cr} + \dif J)$ 
are pushed by the Galactic tide across the boundary of the loss cone. 
The loss cone is filled when $\Delta J \gtrsim J_{\rm cr}$; 
the corresponding injection rate at a given $L$ is \citepalias{HT1986} 
\begin{equation} \label{eq:dif_Gamma_full}
    \Gamma_{\rm f}(L) \, \dif L \simeq \frac{4 \pi^{2} (GM)^{2} J_{\rm cr}^{2}}{L^{3}} f(L,J_{\rm cr}) \, \dif L.
\end{equation} 
This is equal to the steady-state number of comets inside the loss cone ($J \leq J_{\rm cr}$) 
divided by their orbital period $P(L)$. 
Because $\Delta J \propto L^{7} \propto a^{7/2}$, 
the loss cone is empty (filled) for small (large) $L$ or $a$. 
The transition occurs when $\Delta J \simeq J_{\rm cr}$ or, equivalently, the two expressions for $\Gamma(L)$ are equal:
\begin{align}
    a_{\rm eq} &= 0.38 \left( \frac{M^{2} q_{\rm cr}}{\rho_{\rm g}^{2}} \right)^{1/7} \label{eq:afill_tides} \\
    &\approx 2.6 \times 10^{4} \AU \left( \frac{M}{\MSol} \frac{0.1 \MSol \pc^{-3}}{\rho_{\rm g}} \right)^{2/7} \left( \frac{q_{\rm cr}}{1 \AU} \right)^{1/7}. \nonumber
\end{align} 
Figure \ref{fig:losscone_basic} depicts the differences between the empty and filled loss cone regimes schematically.  

We estimate the injection rate as a function of $L$ as
\begin{equation}
    \Gamma(L) = 
    \begin{cases}
        \Gamma_{\rm e}(L), & L < L_{\rm eq}; \\
        \Gamma_{\rm f}(L), & L \geq L_{\rm eq},
    \end{cases}
\end{equation}
where $L_{\rm eq} = (G M a_{\rm eq})^{1/2}$.
We then have 
\begin{equation}
    \Gamma_{\rm tot} = \int_{L_{1}}^{L_{\rm eq}} \Gamma_{\rm e}(L) \, \dif L + \int_{L_{\rm eq}}^{L_{2}} \Gamma_{\rm f}(L) \, \dif L,
\end{equation}
where $L_{1,2} = (G M a_{1,2})^{1/2}$ correspond to the inner and outer edges of the XOC. 
For most realistic distribution functions, 
$\Gamma(L)$ has a peak around $L = L_{\rm eq}$;
thus,
\begin{equation} \label{eq:Gamma_tot_est}
    \Gamma_{\rm tot} \sim L_{\rm eq} \Gamma(L_{\rm eq}) \propto \frac{M^{2} q_{\rm cr}}{a_{\rm eq}} f(L_{\rm eq},J_{\rm cr}).
\end{equation}
We set $a_{2} = 10^{5} \AU$ unless otherwise specified. 
We investigate the effect of varying $a_{1}$ below. 

\begin{figure}
    \centering
    \includegraphics[width=\columnwidth]{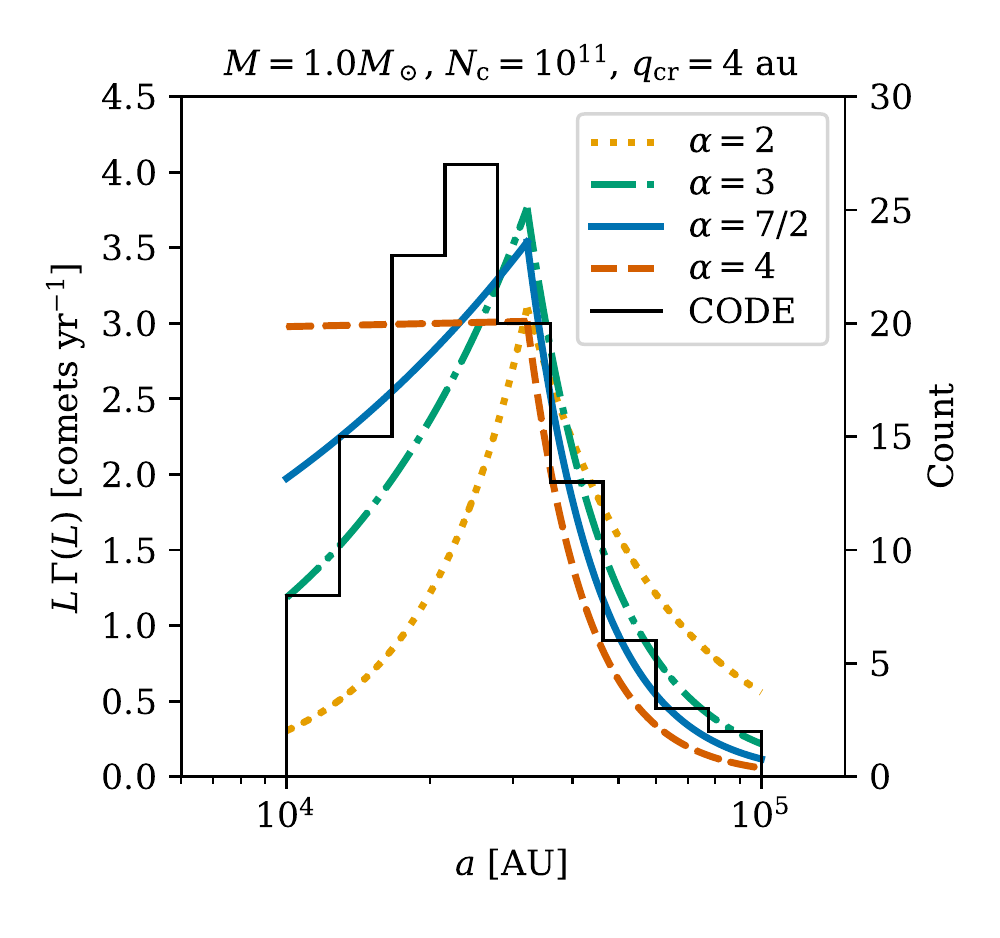}
    \caption{Predicted incremental distribution 
    of dynamically new comets with respect to $a$ for four fiducial OC models (coloured curves), 
    as predicted by the HT86 formalism. 
    All models contain the same number of comets but have different density exponents $\alpha$. 
    The black histogram shows the distribution of 117 observed comets 
    with $a \geq 10^{4} \AU$ and $q \leq 4 \AU$ in CODE \citep{KD2020}.}
    \label{fig:GammaDist_SolSys}
\end{figure}

\subsubsection{Fiducial distribution function} \label{s:OC-Dynamics:Gtide:DistFunc}

In order to evaluate $\Gamma_{\rm tot}$, 
we must specify the distribution function $f(L,J)$. 
Let the number density profile of comets in the XOC be $n(a)$, 
and let the eccentricity distribution be $f_{e}(e)$. 
We assume that $f_{e}$ is independent of semi-major axis.
We then have
\begin{align}
    \dif N &= f(L,J) \, \dif L \, \dif J \int_{-J}^{J} \dif J_{z} \int_{0}^{2 \pi} \dif l \int_{0}^{2 \pi} \dif \chi \int_{0}^{2 \pi} \dif \Omega \nonumber \\
    &= (2\pi)^{3} 2 J f(L,J) \, \dif L \, \dif J \nonumber \\
    &\equiv 4 \pi a^{2} n(a) f_{e}(e) \, \dif a \, \dif e,
\end{align} 
relating $n(a)$ and $f_{e}(e)$ to $f(L,J)$. 
In the following, we assume that comets have a `thermal' eccentricity distribution $f_{e}(e) = 2e$ \citep{Jeans1919, Ambartsumian1937}; 
in this case, $f$ is independent of $J$.

Simulations of OC formation and evolution \citep[e.g.][]{Duncan+1987, Leto+2008, HK2015} 
generally predict a centrally concentrated density profile, 
often approximated as a truncated power law:
\begin{equation} \label{eq:power_law}
    n(a) \propto a^{-\alpha} \hspace{0.5cm} (a_{1} \leq a \leq a_{2})
\end{equation}
We adopt this as our fiducial density profile. 
Thus, our fiducial distribution function is
\begin{align} \label{eq:fdist_fiducial}
    f(L, J) = 
    \begin{cases}
        C N_{\rm c} \left( L/L_{1} \right)^{3-2\alpha}, & L_{1} \leq L \leq L_{2}; \\
        0, & \text{else}.
    \end{cases}
\end{align}
The normalization constant is 
\begin{equation} \label{eq:Cnorm}
    C = 
    \begin{cases}
        (\alpha-3) \left[ 4 \pi L_{1}^{3} \left( 1 - (L_{1}/L_{2})^{2\alpha-6} \right) \right]^{-1},
        & \alpha \neq 3; \\ 
        \left[ 8 \pi^{3} L_{1}^{3} \ln(L_{2}/L_{1}) \right]^{-1}, & \alpha = 3.
    \end{cases}
\end{equation}
The value of $\alpha$ is typically taken to be between $2$ and $4$.

\begin{figure}
    \centering
    \includegraphics[width=\columnwidth]{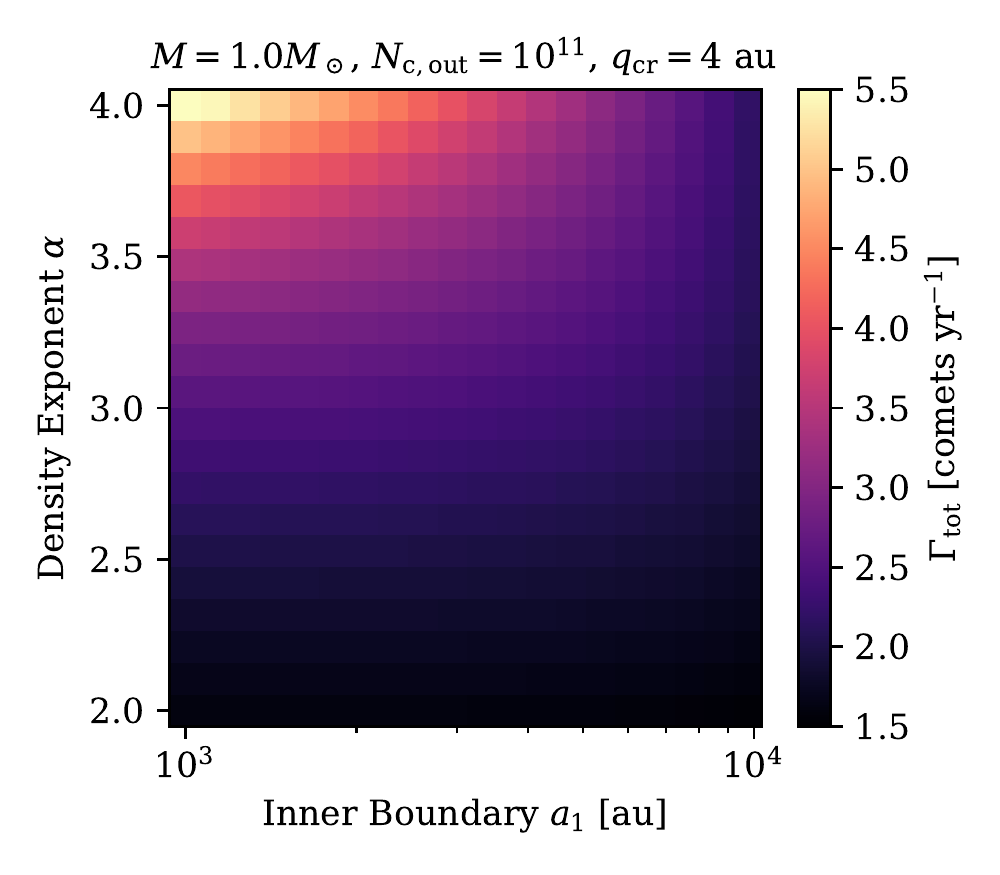}
    \caption{Heat map showing the total LPC flux 
    in OC models with a spherical inner extension (`Hills cloud') 
    as a function of the density power-law exponent $\alpha$ 
    and inner boundary distance $a_{1}$. 
    The number of comets with $a > 10^{4} \AU$ is held fixed.}
    \label{fig:GammaTot_HeatMap}
\end{figure}

\subsubsection{LPC flux in the Solar System}

In this section, we illustrate the validity and limitations of the formalism derived by \citetalias{HT1986}. 
We apply it to estimate the flux of ``dynamically new'' LPCs (defined as those with $a \geq 10^{4} \AU$) 
into the inner Solar System and their incremental distribution with respect to $a$. 
These are given by the quantities $\Gamma_{\rm tot}$ and $L \, \Gamma(L)$. 
We set $q_{\rm cr} = 4 \AU$ (and thus $J_{\rm cr} = [2 G \MSol q_{\rm cr}]^{1/2}$) 
in this exercise, corresponding to $a_{\rm eq} = 3.2 \times 10^{4} \AU$.

We adopt a fiducial OC model containing $N_{\rm c} = 10^{11}$ comets distributed spherically
between $a_{1} = 10^{4} \AU$ and $a_{2} = 10^{5} \AU$ with a power-law exponent in Eq.\ (\ref{eq:power_law}) of  $\alpha = 7/2$. 
This is motivated by theoretical OC formation models \citep{Duncan+1987,VND2019}. 
The predicted total flux is $\Gamma_{\rm tot} = 2.1 \yr^{-1}$, 
with $1.6 \yr^{-1}$ coming from the empty loss cone ($a < a_{\rm eq}$) 
and $0.5 \yr^{-1}$ from the filled loss cone ($a > a_{\rm eq}$). 
This lies at the lower end of the range of observational estimates $\sim 1$--$10 \yr^{-1}$  
\citep[e.g.][]{Everhart1967, Whipple1978, Hughes2001, Francis2005, Bauer+2017, Boe+2019}. 
Figure \ref{fig:GammaDist_SolSys} shows the distribution of new LPCs predicted by the HT86 formalism 
for this model as a solid blue curve. 
To assess the accuracy of the predicted LPC distribution, 
we compare it with a sample of 117 comets with $a \geq 10^{4} \AU$ and $q \leq 4 \AU$
from the Catalogue of Cometary Orbits and their
Dynamical Evolution \citep[CODE; ][]{KD2020}.
The model reproduces the observed distribution moderately well. 
The `Oort peak' of observed comets at $a \approx 2.5 \times 10^{4} \AU$ 
roughly matches the predicted location ($a = a_{\rm eq} \approx 3.2 \times 10^{4} \AU$), 
although the number of comets with $a \lesssim 2 \times 10^{4} \AU$ 
decreases more steeply than the model predicts.

\begin{figure}
    \centering
    \includegraphics[width=\columnwidth]{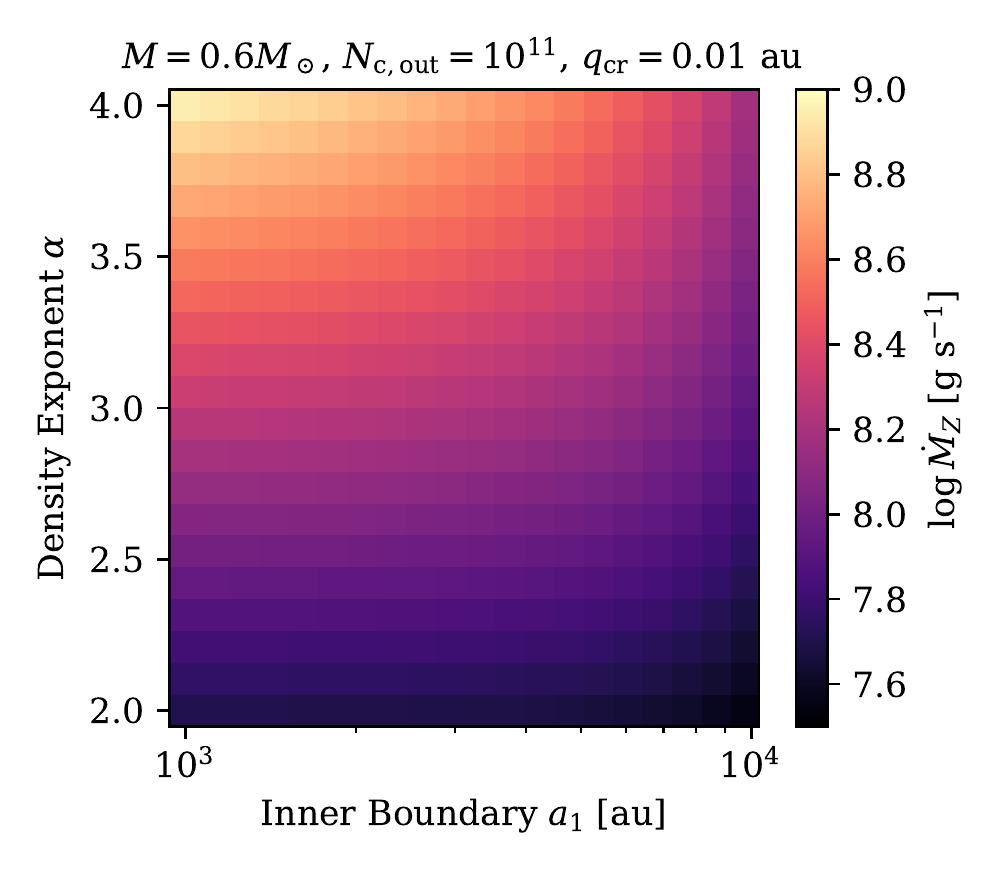}
    \caption{Similar to Fig.\ \ref{fig:GammaTot_HeatMap}, this heat map shows 
    the expected metal accretion rate (as $\log\dot{M}_{Z}$) of a $0.6 \MSol$ WD 
    from XOC comets as a function of $\alpha$ and $a_{1}$. 
    Each accretion event is assumed to contribute a mass $f_{\rm acc} m_{\rm c} = 10^{17} \, {\rm g}$ 
    of metals to the atmosphere.} 
    \label{fig:MdotZ_HeatMap}
\end{figure}

We also show in Fig.\ \ref{fig:GammaDist_SolSys} the predicted LPC distributions 
for several other OC models with the same $N_{\rm c}$ and $a_{1,2}$ but different $\alpha$. 
The predicted total fluxes ($\approx 1$--$2 \yr^{-1}$) are roughly equal for all $\alpha$,
but the incremental distributions vary significantly. 
The relative number of comets with $a < 2 \times 10^{4} \AU$ is somewhat sensitive to $\alpha$. 
The models with $\alpha = 2$ and $4$ are poor matches to the observed distribution, 
but $\alpha = 3$ matches about as well as the $\alpha=7/2$ model. 
As evident from Fig.\ \ref{fig:GammaDist_SolSys}, a single-power-law model of the OC 
cannot perfectly reproduce the observed distribution of dynamically new comets.
However, models with $\alpha \approx 3$--$3.5$ are adequate to reproduce the broad features. 
For the remainder of this work, we adopt $\alpha = 7/2$ unless otherwise specified.

The `inner' OC ($a < 10^{4} \AU$, a.k.a.\ the Hills cloud) 
was not considered an important source of LPCs 
until relatively recently \citep[e.g.][]{KQ2009, DK2011}. 
Comet orbits in this region are dynamically stabler 
because the Galactic tide is weaker ($\dot{J} \propto a^{2}$, Eq.\ \ref{eq:dJdt}). 
The Jovian planets affect inner OC comets more strongly 
as they approach the inner Solar System, 
sometimes making them appear to originate from the outer OC \citep{KQ2009}.
Meanwhile, theoretical studies of OC formation and evolution have predicted 
a variety of cloud populations and density profiles 
interior to $10^{4} \AU$ \citep[e.g.][]{Duncan+1987, KQ2008, VND2019}. 
The properties of the inner OC are therefore highly uncertain, 
but potentially significant.

Faced with these uncertainties, it is instructive to see how the predicted LPC flux changes 
when we include a spherically symmetric, dynamically relaxed inner OC in our model.
We therefore repeat the previous exercise for a variety of OC models 
in which the density profile is extended inward to an inner boundary distance $a_{1}$, 
keeping the number of comets with $a > 10^{4} \AU$ constant. 
We show the predicted comet flux $\Gamma_{\rm tot}$ in Figure \ref{fig:GammaTot_HeatMap} 
for a grid of $a_{1}$ and $\alpha$ values. 
The value of $\Gamma_{\rm tot}$ is more sensitive to $a_{1}$ for larger $\alpha$, 
i.e.\ for more centrally concentrated density profiles.
The LPC flux varies by a factor of $\approx 4$ over this section of parameter space, 
never exceeding $\approx 6 \yr^{-1}$. 
Thus, the presence of an inner OC increases the potential LPC flux only by a factor of a few. 

Overall, the HT86 formalism, coupled to a spherical, dynamically relaxed OC model with a power-law density profile, 
provides a good estimate of the LPC flux in the Solar System, 
although some properties of the OC are relatively unconstrained. 
The formalism also broadly reproduces the observed distribution of semi-major axes. 

\subsubsection{WD pollution from XOC comets} \label{s:OC-Dynamics:Gtide:Flux-at-WD}

In this subsection, we estimate the metal accretion rate of a WD 
due to the tidal disruption of LPCs with small periastron distances. 
We apply analogous calculations as in the previous subsection, 
assuming a typical WD with mass $M \approx 0.6 \MSol$.
We assume that the comets are rubble piles 
with uniform mass density $\rho_{\rm c}$ and zero internal strength. 
Therefore, the comets are tidally disrupted when they pass closer within a distance $r_{\rm tide}$ of the WD, 
where
\begin{align} \label{eq:TidalDisruption}
    r_{\rm tide} &\simeq \left( \frac{M}{\rho_{\rm c}} \right)^{1/3} \\
    &\approx 8 \times 10^{-3} \AU \left( \frac{M}{0.6 \MSol} \right)^{1/3} \left( \frac{\rho_{\rm c}}{0.6 \, {\rm g \, cm^{-3}}} \right)^{-1/3}. \nonumber
\end{align}
For simplicity, we assume that there are no planets orbiting the WD massive enough to perturb the orbits of LPCs. 
Although comets pass close to the WD, short-range dynamical effects such as general relativistic precession 
have a negligible effect on the rate at which comets enter the loss cone. 
This is because the GT torque is applied mainly near apoastron.
The HT86 formalism can therefore be used to estimate the tidal disruption rate $\Gamma_{\rm dis}$. 
This disruption rate is equal to the value of $\Gamma_{\rm tot}$ when $q_{\rm cr} = r_{\rm tide}$ ($J_{\rm cr} = [2 G M r_{\rm tide}]^{1/2} \equiv J_{\rm tide}$). 
The disruption rate is related in turn to the rate $\dot{M}_{Z}$ 
at which metal-rich debris pollutes the WD atmosphere. 
If the average mass of a comet is $m_{\rm c}$ 
and a fraction $f_{\rm acc}$ of the debris from each comet is eventually accreted by the WD, then
\begin{equation}
    \dot{M}_{Z} = f_{\rm acc} m_{\rm c} \Gamma_{\rm dis}.
\end{equation}
We adopt a representative value $m_{\rm c} = 10^{17} \, {\rm g}$, 
corresponding to a total OC mass $M_{\rm OC} = m_{\rm c} N_{\rm c} = 1.7 \ME (N_{\rm c} / 10^{11})$
\citep[cf.][]{Weissman1996, Francis2005, Boe+2019}. 
For a spherical body of density $\rho_{\rm c} = 0.6 \, {\rm g \, cm^{-3}}$, 
$m_{\rm c} = 10^{17} \, {\rm g}$ corresponds to a radius of $3.4 \, {\rm km}$.

We assume that single WDs have XOCs with broadly similar properties to the OC in the Solar System.
Adopting a representative value $q_{\rm cr} = r_{\rm tide} \approx 0.01 \AU$, we find $a_{\rm eq} = 1.2 \times 10^{4} \AU$. 
It is straightfoward to calculate $\dot{M}_{Z}$ for the family of models from the previous section. 
Figure \ref{fig:MdotZ_HeatMap} shows the result in an analogous manner to Fig.\ \ref{fig:GammaTot_HeatMap}. 
We find
\begin{equation} \label{eq:MdotZ_naive}
    \dot{M}_{Z} \sim 10^{8 \pm 1} \, {\rm g \, s^{-1}} \left( \frac{f_{\rm acc} m_{\rm c}}{10^{17} \, {\rm g}} \right) \left( \frac{N_{\rm c,out}}{10^{11}} \right),
\end{equation}
where $N_{\rm c,out}$ is the number of comets with $a > 10^{4} \AU$.
Note that the magnitude of $\dot{M}_{Z}$ is much more sensitive to $a_{1}$ and $\alpha$ in this case 
because of the smaller value of $q_{\rm cr}$ (and hence $a_{\rm eq}$). 
This shows the importance of accounting for the possible existence of an inner OC around a WD, 
even if this region does not necessarily dominate the LPC flux in the Solar System.

\subsection{Stellar flybys} \label{s:OC-Dynamics:flybys}

\subsubsection{Weak encounters}

Distant stellar flybys induce weak stochastic perturbations on OC comets. 
These cause a phase-space diffusion which smooths the distribution function 
and injects new comets with small $q$.  
As a first approximation, comet injection by flybys 
may be treated as a separate process from injection by the GT \citep[but see][]{CS2010}. 

\citetalias{HT1986} showed the comet injection rate 
due to weak flybys has a similar behaviour to the rate due to the GT.
As before, there are distinct regimes where the loss cone is empty and filled. 
The transition distance between regimes is modified by the fact 
that only stars contribute to the density (cf.\ Eq.\ \ref{eq:afill_tides}):
\begin{equation} \label{eq:afill_flyby}
    a_{\rm eq*} = 0.32 \left( \frac{M^{2} q_{\rm cr}}{\rho_{*}^{2}} \right)^{1/7},
\end{equation}
where $\rho_{*}$ is the mean stellar density of the Galactic environment. 
The differential and total injection rates due to stellar flybys, $\Gamma_{*}(L)$ and $\Gamma_{\rm tot*}$, 
have the same essential scalings as in the GT problem. 
By Eq.\ (\ref{eq:Gamma_tot_est}), the ratio of the comet injection rates 
from each mechanism is
\begin{equation}
    \frac{\Gamma_{\rm tot*}}{\Gamma_{\rm tot}} \sim \frac{a_{\rm eq}}{a_{\rm eq*}} \frac{f(L_{\rm eq*}, J_{\rm cr})}{f(L_{\rm eq}, J_{\rm cr})} = \left( \frac{\rho_{*}}{\rho_{\rm g}} \right)^{(2 \alpha - 1)/7}.
\end{equation}
The stellar mass density of the Galactic disc, $\rho_{*}$, is somewhat less than $\rho_{\rm g}$; 
for example, in the solar neighbourhood, $\rho_{*} = 0.04 \MSol \pc^{-3}$ 
versus $\rho_{\rm g} = 0.1 \MSol \pc^{-3}$ \citep[e.g.][]{Flynn+2006, McKee+2015}.
Therefore, we expect $\Gamma_{\rm tot*} / \Gamma_{\rm tot} \lesssim 1$ for a realistic $\alpha$.

\citetalias{HT1986} argued on this basis that GTs are the dominant mechanism 
by which LPCs are injected into the inner Solar System (except during comet showers; see below). 
The same reasoning applies to comets orbiting WDs in the Galactic disc. 
The metal accretion rate derived from GTs and weak stellar flybys together 
is therefore approximately the same as that due to GTs alone. 

\subsubsection{Comet showers} \label{s:OC-Dynamics:showers}

Field stars occasionally pass through the inner portion of the Oort Cloud, triggering comet showers. 
During these events, the comet injection rate is enhanced by at least an order of magnitude \citep*{Hills1981,HTA1987}. 
If XOCs are common around WDs, 
then comet showers could elevate the rate of metal accretion 
over the `background' rate due to Galactic tides and weak flybys.

Comet showers are relatively infrequent events. 
The duration $\Delta t_{\rm close} = b/v_{*}$ of a close encounter with a field star with impact parameter $b$ and speed $v_{*}$ 
is short compared to the orbital period of a comet at the same distance. 
Therefore, the duration of each shower is of the order of $P$.
For a typical $v_{*}$, we use the 3D velocity dispersion of thin-disc stars from \citet{Anguiano+2020}. 
This gives
\begin{align}
    \Delta t_{\rm close} &\approx 10^{-3} \Myr \left( \frac{b}{10^{4} \AU} \right) \left( \frac{v_{*}}{50 \, {\rm km \, s^{-1}}} \right)^{-1}, \\
    P &\approx 1.3 \Myr \left( \frac{b}{10^{4} \AU} \right)^{3/2} \left( \frac{M}{0.6 \MSol} \right)^{-1/2}.
\end{align}
Meanwhile, the expected interval between flybys 
with impact parameter closer than a distance $b$ is
\begin{align}
    \Delta t_{\rm flyby} &= \frac{1}{\pi b^{2} n_{*} v_{*}} \nonumber \\
    &\approx 30 \Myr \left( \frac{b}{10^{4} \AU} \right)^{-2} \left( \frac{n_{*}}{0.1 \pc^{-3}} \right)^{-1} \left( \frac{v_{*}}{50 \, {\rm km \, s^{-1}}} \right)^{-1}, \label{eq:flyby_rate}
\end{align}
where $n_{*}$ is the number density of field stars. 
In practice, only stars more massive than the WD can trigger comet showers \citep{HTA1987}. 
This effectively reduces $n_{*}$ from the nominal value 
by a factor related to the stellar mass function, 
for the purposes of determining the frequency of comet showers. 
Using a nominal WD mass $M = 0.6 \MSol$, 
and assuming that passing stars are drawn from the present-day mass function fitted by \citet{Bovy2017}, 
we estimate the fraction of close flybys by stars 
more massive than $M$ to be $f_{M} \approx 0.14$.
If comet showers occur for $b \leq 10^{4} \AU$, 
then their `duty cycle' is
\begin{align}
    p_{\rm sh} &= \frac{f_{M} P}{\Delta t_{\rm flyby}} \nonumber \\
    &\approx 6 \times 10^{-3} \left( \frac{b}{10^{4} \AU} \right)^{7/2} \left( \frac{f_{M} n_{*}}{0.014 \pc^{-3}} \frac{v_{*}}{50 \, {\rm km \, s^{-1}}} \right).
\end{align}
A comet-hosting WD therefore has a $\sim 1$ per cent probability 
of being in the midst of a comet shower when we observe it. 
In a sample of $\mathcal{N}$ single WDs that possess XOCs, 
the probability that a comet shower is ongoing in one system is
\begin{equation}
    \mathcal{N} p_{\rm sh} (1 - p_{\rm sh})^{\mathcal{N}-1}.
\end{equation}
Detailed spectroscopic analyses that constrain the presence of volatiles
have been published for a few dozen WDs at present \citep[e.g.][]{KGF2014}. 
For a nominal value $p_{\rm sh} = 0.006$, 
we calculate a $\approx 15$ per cent probability of an ongoing comet shower 
in a sample of size $\mathcal{N} = 30$. 
Thus, the current subset of polluted WDs with constrained volatile abundances  
may not be large enough to constrain the occurrence of comet showers.

\subsection{Synthesis}

Our model predicts that, if single WDs possess XOCs with properties broadly similar to the Solar System's OC, 
a large fraction would be polluted by volatile-rich debris 
with characteristic accretion rates $\dot{M}_{Z} \sim 10^{8 \pm 1} \, {\rm g \, s^{-1}}$ (see Eq.\ \ref{eq:MdotZ_naive}). 
The predicted $\dot{M}_{Z}$ is consistent with the median inferred accretion rate among polluted WDs, 
which is around $10^{8} \, {\rm g \, s^{-1}}$ \citep[e.g.][]{Wyatt+2014, Xu+2019, BX2022}. 
The minimum accretion rate required for pollution 
to be detected in a typical WD atmosphere is $\sim 10^{5} \, {\rm g \, s^{-1}}$ \citep[e.g.][]{KGF2014, BX2022}. 
The predicted pollution would therefore be detectable by current observations, 
even if the accretion process is fairly inefficient ($f_{\rm acc} \sim 0.01$--$0.1$). 
It is difficult to explain the paucity of polluted WD atmospheres with a comet-like composition 
within the standard theory of comet delivery from an XOC.

There are three main possibilities to explain this apparent discrepancy:
\begin{enumerate}
    \item XOCs are intrinsically rare around WD progenitors. 
    \item XOCs are depleted during late stellar evolution. 
    \item Additional dynamical processes interfere with the delivery of XOC comets 
    to extremely small periastron distances ($q \sim r_{\rm tide}$).
\end{enumerate}
Option (i) has not been ruled out by direct observations. 
However, it would be difficult to reconcile this scenario with 
the longstanding idea that XOC formation is an expected byproduct 
of the formation and early dynamical evolution of giant planets \citep[e.g.][]{Duncan+1987, HM1999, VND2019}. 
The progenitors of polluted WDs are more massive than the Sun on average \citep[e.g.][]{KGF2014}, 
and the occurrence rate of giant planets has a positive correlation with stellar mass 
\citep[e.g.][]{Johnson+2010, Reffert+2015, Jones+2016, Ghezzi+2018}. 
If indeed giant-planet formation typically results in XOC formation, 
then WD progenitors may be more likely to possess XOCs than solar-mass stars. 
The Galactic population of interstellar interlopers 
may also be consistent with ubiquitous XOC formation 
(\citealt{Do+2018}; but see \citealt{MoroMartin2018, MoroMartin2019}). 
In the next two sections, therefore, we consider two mechanisms that could plausibly contribute to options (ii) and (iii). 

\section{Stellar Mass Loss} \label{s:MassLoss}

The effects of stellar evolution on OC comets are most pronounced 
during the asymptotic giant branch (AGB) phase, 
when the star ejects a large fraction of its mass in a wind. 
Mass loss causes cometary orbits to expand and possibly become unbound \citep{Veras+2011}. 
If AGB mass loss is somewhat asymmetric, a natal kick is imparted to the WD. 
The typical time-scale of AGB mass loss ($\sim 0.1$--$1 \Myr$) and WD kick speed ($\sim 1 \, {\rm km \, s^{-1}}$; e.g.\ \citealt{Heyl2007,Heyl2008,Heyl2008b}; \citealt{Davis+2008}; \citealt{HP2009}; \citealt{Fregeau+2009}; \citealt{Izzard+2010}; \citealt{EBR2018}) 
are comparable to the orbital period and velocity of OC objects. 
These effects will reduce the number of comets in the cloud 
and may alter their orbital distribution function. 
In this section, we evaluate to what extent these processes reduce the WD comet accretion rate. 
In subsection \ref{s:MassLoss:Orbits}, we describe our parametric model for stellar mass loss.
In subsection \ref{s:MassLoss:Numerics}, we conduct a series of numerical integrations of XOC objects 
in the context of this model. 
Previous studies along similar lines include \citet{PA1998}, \citet{Veras+2014}, \citet{SML2015}, and \citet{CH2017}.
We compare our methods and results to theirs in Section \ref{s:Discussion:Compare}.

\subsection{Asymmetric mass loss} \label{s:MassLoss:Orbits}

Consider a star with mass $M$ steadily ejecting material at a rate $\dot{M}$. 
We assume that the outflow exhibits a moderate asymmetry, 
such that the star receives a recoil acceleration in a fixed direction (unit vector $\khat$). 
This acceleration is given by
\begin{equation}
    \vect{g}_{*} = \frac{u_{\rm e} \dot{M}}{M} \khat,
\end{equation}
where $u_{\rm e}$ is an effective `exhaust speed' determined by the velocity and geometry of the outflow. 
The final velocity of the star with respect to its initial rest frame is
\begin{equation}\label{eq:velocity_kick}
    \vect{V}_{\rm k} = u_{\rm e} \ln\left( \frac{M_{\rm i}}{M_{\rm f}} \right) \khat\,.
\end{equation}
In Eq.\ (\ref{eq:velocity_kick}), $M_{\rm i}$ and $M_{\rm f}$ are the initial and final mass of the star, respectively, 
which are given roughly by the zero-age main sequence mass ($M_{\rm i} \equiv M_{\rm ms}$) 
and the WD mass ($M_{\rm f} \equiv M_{\rm wd}$). 
We adopt fiducial values $M_{\rm ms} = 2.0 \MSol$ and $M_{\rm wd} = 0.56 \MSol$ 
based on the theoretical initial--final mass relation of \citet{Choi+2016}.

We assume that the stellar mass decreases exponentially after $t = 0$ according to
\begin{equation}
    M(t) = M_{\rm ms}  \left( \frac{M_{\rm ms}}{M_{\rm wd}} \right)^{-t/T}\,,
\end{equation}
until it reaches the final mass $M_{\rm wd}$ at $t = T$. 
This model has the helpful property that $M / \dot{M}$ is constant, 
meaning that $g_{*}$ is constant as well. 

This parametric mass-loss model is admittedly idealized. 
In detailed evolutionary models, late-stage mass loss takes place in `bursts' coinciding with thermal pulses on the AGB. 
To complicate the situation further, the recoil acceleration may not have a fixed direction throughout mass loss. 
This symmetry breaking could change the resulting dynamics of injected comets.

\subsection{Numerical setup} \label{s:MassLoss:Numerics}

We perform simulations of comets orbiting a star undergoing continuous mass loss with {\sc rebound} \citep{RL2012}. 
The Newtonian equations of motion are solved in the rest frame of the central star.
We include the effects of stellar mass loss via
the {\tt modify\_mass} operator in the extension package {\sc reboundx} \citep{Kostov+2016,Tamayo+2020}. 
This operator causes the mass of a particle -- in this case the central star -- 
to grow or decay exponentially on a user-defined time-scale as described above. 
Mass loss in {\sc reboundx} is assumed to be isotropic. 
To simulate the effects of asymmetric mass loss, 
we include an additional fictitious force per unit mass acting on the comet,
defined to be equal and opposite to the recoil acceleration 
of the central star with respect to an inertial frame. 

The goals of our simulations are as follows:
\begin{enumerate}
    \item[(i)] To evaluate the fraction of OC comets that 
    remain bound to a star after AGB mass loss and a kick.
    \item[(ii)] To examine the change 
    of the distribution function of comets as a result of these processes.
    \item[(iii)] To test how varying the mass-loss time-scale 
    and the magnitude of the kick imparted to the star affects these outcomes.
\end{enumerate}

\begin{figure}
    \centering
    \includegraphics[width=\columnwidth]{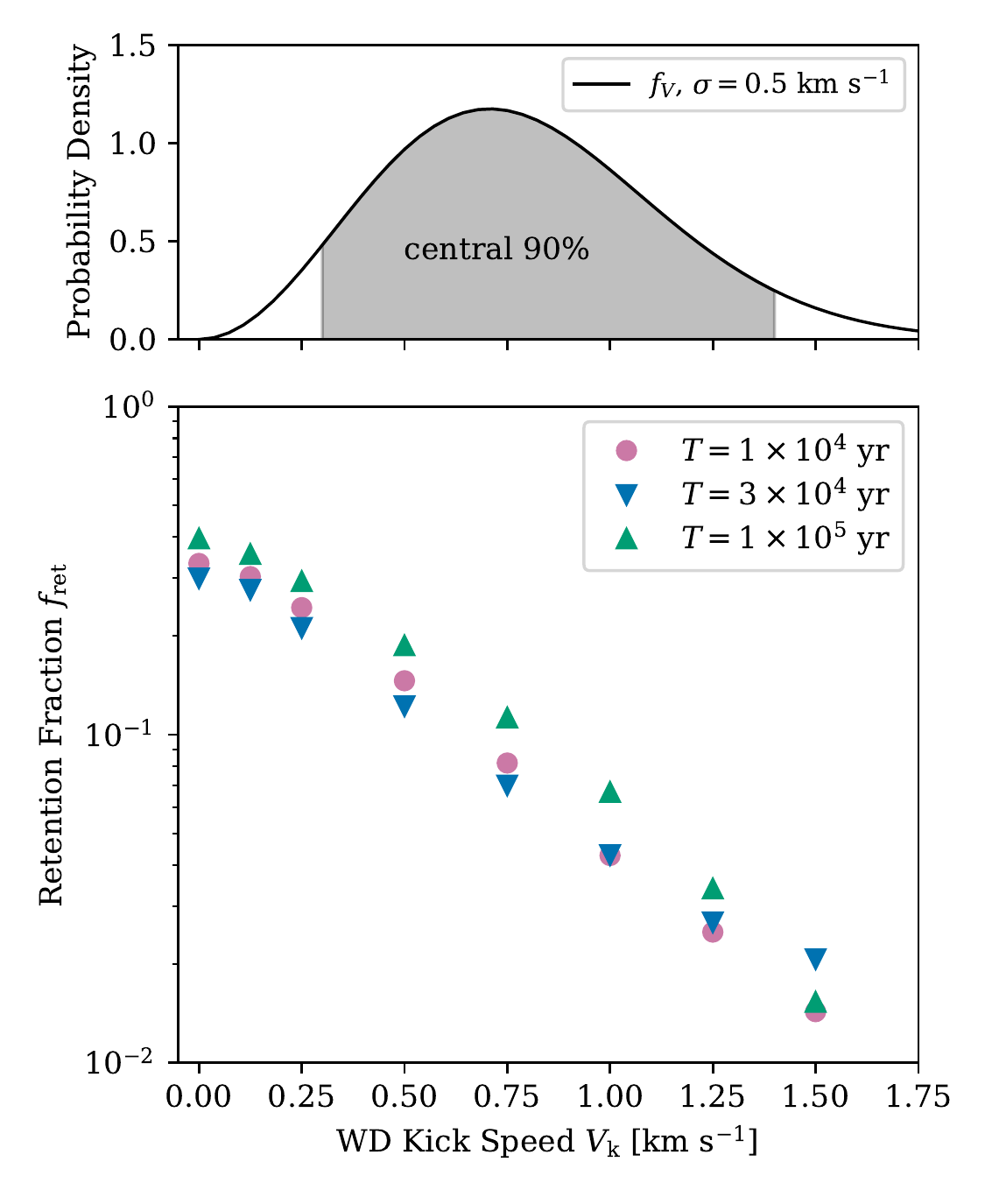}
    \caption{{\it Lower panel:} Fraction of comets retained in bound orbits 
    as a function of the kick speed $V_{\rm k}$ imparted to the WD (horizontal axis) 
    and the stellar mass-loss time-scale (different shapes and colours). 
    Each point represents an ensemble of $10^{4}$ simulations. 
    {\it Upper:} The black curve shows a Maxwellian distribution of kick speeds (Eq.\ \ref{eq:Maxwellian_fV})
    with $\sigma_{\rm k} = 0.5 \, {\rm km \, s^{-1}}$. 
    The shaded region indicates the central 90 per cent quantile range.}
    \label{fig:frac_surv_kick}
\end{figure}

\subsubsection{Initial conditions and parameters}  \label{s:MassLoss:Numerics:InitCond}

We initialize all of our simulations with a $2.0 \MSol$ star 
at rest at the origin and a single comet in orbit. 
We generate the astrocentric orbital elements of each comet as follows: 
The semi-major axis is drawn from a power-law radial density profile 
with $\alpha = 7/2$ between $10^{3}$ and $10^{5} \AU$, 
and the other elements are drawn from their respective isotropic distributions 
(uniform in $e^{2}$, $\cos{I}$, $\chi$, $\Omega$, and $l$). 
Note that we define the inclination $I$ relative to the axis of the WD kick 
in this discussion.

The time-scale of mass loss and the magnitude of the recoil acceleration are chosen such that, 
over the duration $T$ of each simulation, the central star evolves to a final mass of $0.56 \MSol$ 
and attains a given speed $V_{\rm k}$ with respect to its original rest frame. 
We conduct simulations using the following values of these parameters: 
\begin{align*}
    T &= \{ 1 \times 10^{4}, 3 \times 10^{4}, 1 \times 10^{5} \} \yr, \\
    V_{\rm k} &= \{ 0, 0.125, 0.25, 0.50, 0.75, 1.00, 1.25, 1.50 \} \, {\rm km \, s^{-1}}.
\end{align*}
These values of $T$ are representative of the range of time-scales on which WD kicks are thought to be imparted. 
However, these timescales are somewhat shorter than the overall duration of AGB mass loss. 
We therefore tend to over-estimate the magnitude of mass-loss-related effects on an XOC.
The values of $V_{\rm k}$ are based on the dispersion of WD kick speeds inferred by \citet{EBR2018} 
from the occurrence rates of wide binaries containing WDs observed by {\it Gaia}. 

\subsubsection{Procedure and post-processing} \label{s:MassLoss:Numerics:Procedure}

For each combination ($T, V_{\rm k}$), we conduct $10^{4}$ single-comet integrations in {\sc rebound}. 
We use the IAS15 integrator \citep{RS2015} with a sufficiently small time-step 
to resolve the periastron passage of each comet. 

At the end of each integration, we compute the new Keplerian orbit of each comet. 
The fate of each comet -- either ejection or retention -- is prescribed as follows.
We consider a comet ejected if (a) its final orbit is open ($e \geq 1$) 
or (b) its final orbit is closed ($e < 1$) but has an apoastron distance $Q = a (1 + e)$ 
greater than the Hill radius of a WD in the solar neighbourhood (about $0.8 \pc$). 
Comets meeting condition (b) are bound to the newborn WD in the context of the 2-body problem 
but in practice are likely to be stripped by the Galactic tide or field stars. 
Comets that do not satisfy  (a) or (b) are considered to be retained by the WD.

A fraction of the ``ejected'' comets are on incoming orbits with small $q$. 
These comets may experience one final close approach to the WD before ejection.
It is possible that some of these objects are tidally disrupted and accreted by the newborn WD \citep[see][]{SML2015}. 
However, these comets will still be ejected within an orbital period; 
they are not relevant for the long term pollution of the WD.

\subsection{Numerical results} \label{s:MassLoss:Results}

\subsubsection{Retention of OC comets} \label{s:MassLoss:Results:Retention}

Figure \ref{fig:frac_surv_kick} shows the fraction of comets retained in our simulations, $f_{\rm ret}$. 
This fraction decreases steadily with increasing kick speed, 
since the chosen non-zero values of $V_{\rm k}$ are comparable to,
or larger than, the escape speed near the inner edge of the cloud. 
For a fixed $V_{\rm k}$, $f_{\rm ret}$ changes by less than a factor of 2
for different $T$, at least for the range of values we tested. 

To gauge the range of likely values for $f_{\rm ret}$ in real systems, 
we consider a Maxwellian distribution for the WD kick speeds 
with dispersion $\sigma = 0.5 \, {\rm km \, s^{-1}}$, 
with a probability density function 
\begin{equation} \label{eq:Maxwellian_fV}
    f_{V}(V_{\rm k}) = \sqrt{\frac{2}{\pi}} \frac{V_{\rm k}^{2}}{\sigma^{3}} \exp\left( - \frac{V_{\rm k}^{2}}{2 \sigma^{2}} \right).
\end{equation}
The same distribution was used by \citet{EBR2018} to explain 
the separation distribution of wide binaries containing WDs.
We show this distribution in the upper panel of Fig.\ \ref{fig:frac_surv_kick}. 
The mode of the distribution is $V_{\rm k} = \sqrt{2} \sigma \approx 0.7 \, {\rm km \, s^{-1}}$, 
which gives a retention fraction of $\approx 0.1$. 
About $90$ per cent of newborn WDs would retain 
a fraction $0.03 \leq f_{\rm ret} \leq 0.3$ of their OC comets. 
This suggests that the rate of WD comet accretion due to Galactic tides 
would be reduced compared to the estimate in Section \ref{s:OC-Dynamics:Gtide:Flux-at-WD} 
by a factor of $\sim 3$--$30$ with all else equal. 
However, we have not yet taken into account the change of the distribution function of the surviving comets.

\begin{figure}
    \centering
    \includegraphics[width=\columnwidth]{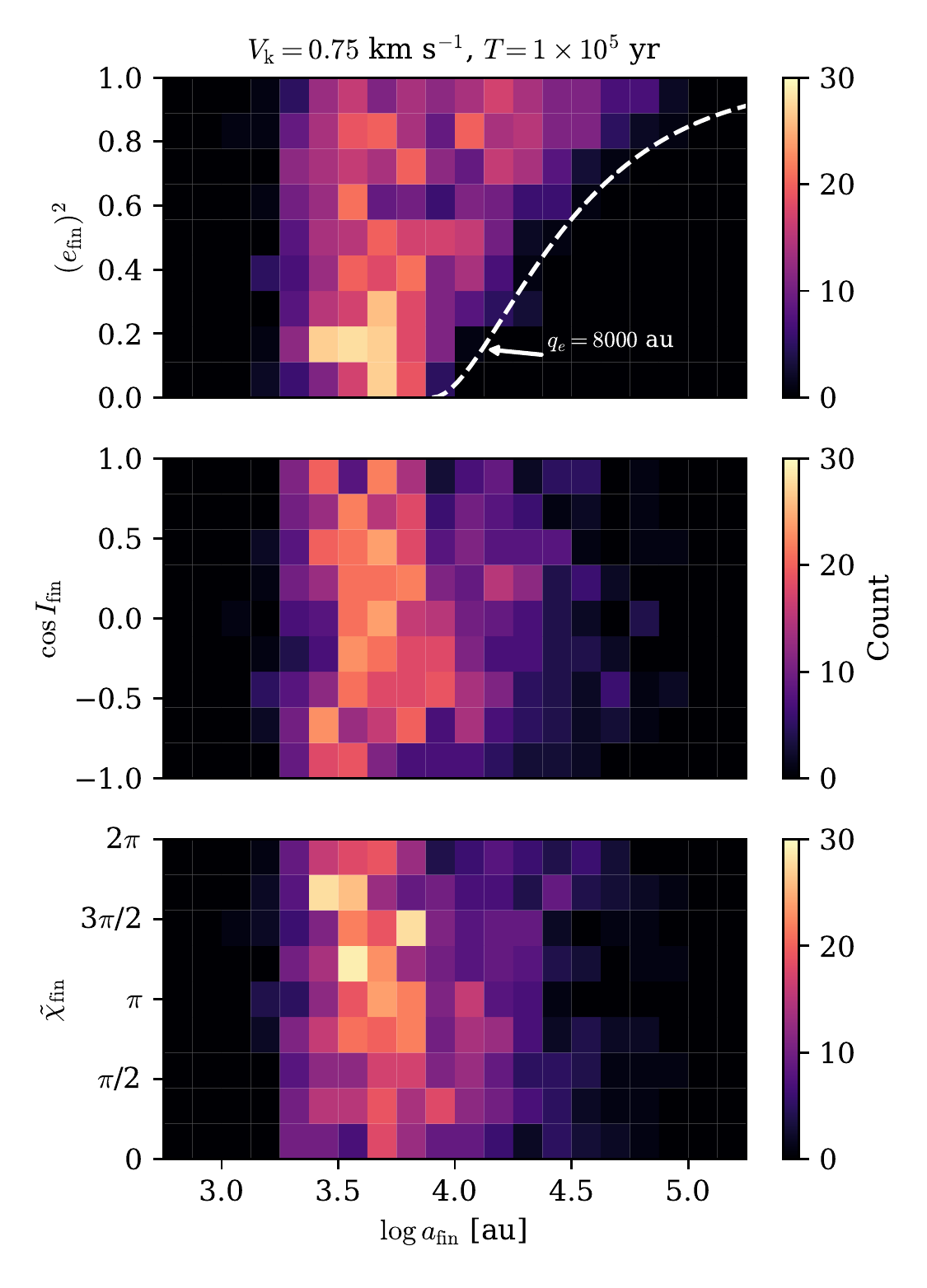}
    \caption{{\it Top panel:} 2D histogram of retained comets in the $(a, e^{2})$ plane 
    for $V_{\rm k} = 0.75 \, {\rm km \, s^{-1}}$ and $T = 1 \times 10^{5} \yr$. 
    The dashed black curve is a curve of constant $q = 8000 \AU$. 
    {\it Middle:} The same in the $(a, \cos{I})$ plane. 
    {\it Bottom:} The same in the $(a, \tilde{\chi})$ plane.}
    \label{fig:PhiAfterKick_V750_T1e5}
\end{figure}

\begin{figure}
    \centering
    \includegraphics[width=\columnwidth]{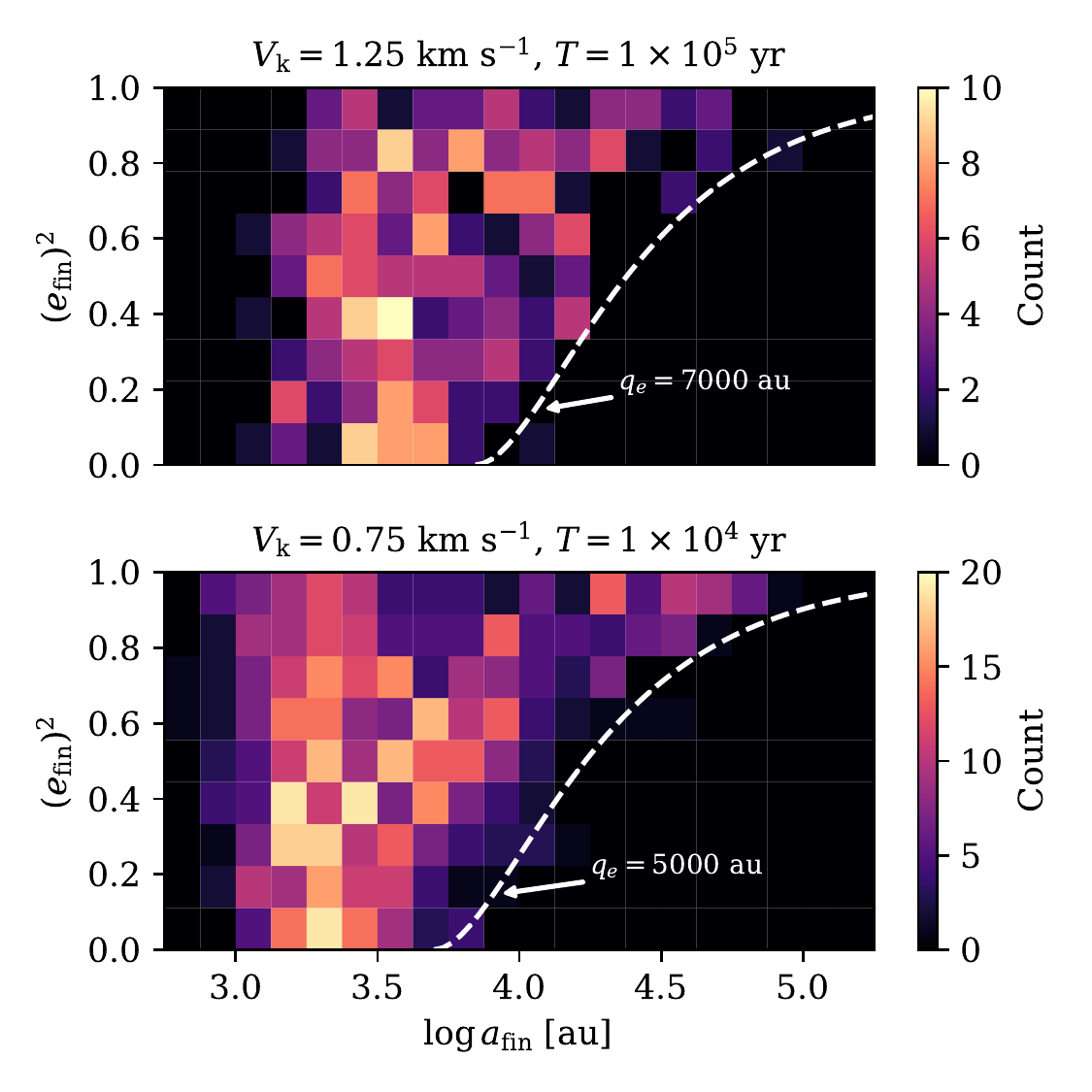}
    \caption{{\it Top panel:} The same as the top panel of Fig.\ \ref{fig:PhiAfterKick_V750_T1e5}, 
    but for $V_{\rm k} = 1.25 \, {\rm km \, s^{-1}}$. 
    {\it Bottom:} The same, but for $V_{\rm k} = 0.75 \, {\rm km \, s^{-1}}$ and $T = 1 \times 10^{4} \yr$.}
    \label{fig:PhiAfterKick_V750T1e5-V750T1e4}
\end{figure}

\subsubsection{Distribution function of retained comets} \label{s:MassLoss:Results:DistFunc}

In order to use the formalism of \citetalias{HT1986} to estimate 
the rate of comet injection after stellar evolution, 
we need to characterize the distribution function $f(L,J,\cdots)$ 
of surviving comets near the edge of the loss cone ($J \ll L$). 
By construction, the distribution function of surviving comets is uniform with respect to $\Omega$ 
if this angle is measured in the plane normal to the recoil acceleration. 
The surviving comets will exhibit a highly non-uniform distribution with respect to $l$ 
because comets near periastron ($l \approx 0, 2\pi$) 
during late stellar evolution are preferentially retained. 
However, this distribution will be isotropized over many orbits. 
We therefore proceed assuming a uniform distribution of $l$ when calculating 
the injection rate. 
For convenience, we define a modified apsidal angle 
\begin{equation}
    \tilde{\chi} \equiv 2 \chi \bmod 2 \pi,
\end{equation}
so that comets that are pushed inward by the Galactic tide ($\dot{J} < 0$) have $0 < \tilde{\chi} < \pi$ (see Eq.\ \ref{eq:dchidt}). 

In Figure \ref{fig:PhiAfterKick_V750_T1e5}, we show the distribution of the surviving comets 
in the planes ($a_{\rm fin}, e_{\rm fin}^{2}$), ($a_{\rm fin}, \cos{I_{\rm fin}}$), and ($a_{\rm fin}, \tilde{\chi}_{\rm fin}$) 
for the nominal case $V_{\rm k} = 0.75 \, {\rm km \, s^{-1}}$ and $T = 1 \times 10^{5} \yr$. 
The subscript `fin' indicates the value at the end of a simulation. 
Figure \ref{fig:PhiAfterKick_V750T1e5-V750T1e4} shows the distribitions in ($a_{\rm fin}, e_{\rm fin}^{2}$) 
for two other combinations of $V_{\rm k}$ and $T$; 
the distributions of $\cos{I_{\rm fin}}$ and $\tilde{\chi}_{\rm fin}$ 
are essentially the same as in the nominal case. 

The retained comets with relatively small $a$ remain well described by a thermal eccentricity distribution; 
that is, at a given $a$, they are uniformly distributed in $e^{2}$, 
with perhaps a moderate enhancement of the number of comets with low eccentricities. 
For $a \gtrsim 10^{4} \AU$, only objects with high eccentricities are retained, 
and the minimum eccentricity increases with $a$. 
This is because objects at large distances are more likely to survive 
if the kick occurs whilst they are near periastron. 
For a given experiment, the ``edge'' of the survivors' distribution 
can be approximated as a curve of constant $q \equiv q_{\rm e}$, 
the location of which may be somewhat sensitive to $V_{\rm k}$ and $T$. 
For $q \leq q_{\rm e}$, the eccentricities of comets follow a truncated thermal distribution. 
The inclinations and apsidal angles of retained comets 
are isotropically distributed 
(uniform in $\cos{I}$ and $\tilde{\chi}$) at all $a$. 

\begin{figure}
    \centering
    \includegraphics[width=\columnwidth]{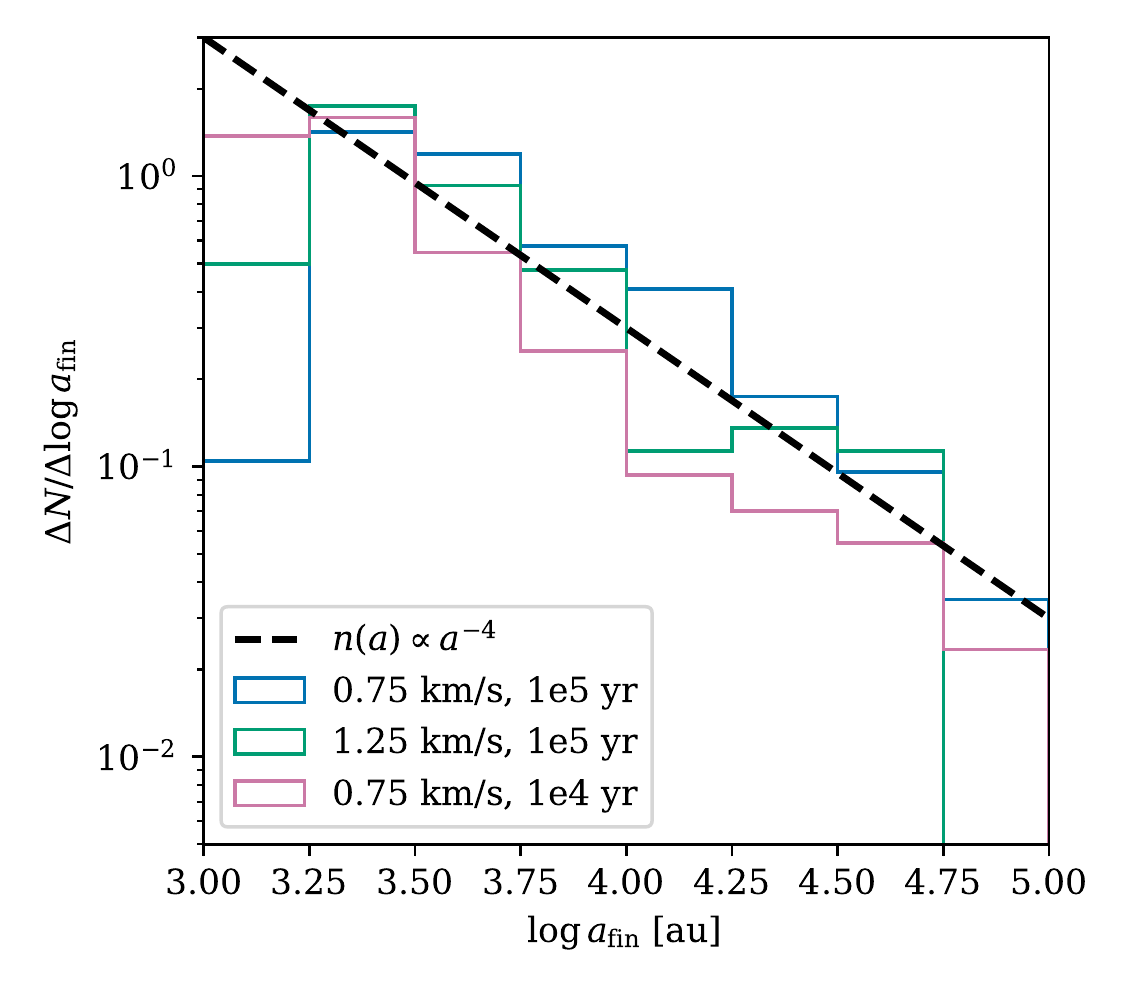}
    \caption{Incremental distributions of the final semi-major axis of retained comets 
    with final $q \leq 1000 \AU$ 
    for different $V_{\rm k}$ and $T$ values. 
    The dashed line indicates a power-law density profile $n(a) \propto a^{-4}$.}
    \label{fig:PostKickPowerLaw}
\end{figure}

The initial cometary orbits followed a power-law density profile $n(a) \propto a^{-\alpha}$ with $\alpha = 7/2$. 
The preferential retention of comets with small $q$ 
implies that the distribution function of retained comets $f(L,J)$ 
is not a pure power-law in $L$ (or $a$). 
However, from Eqs.\ (\ref{eq:dif_Gamma_empty}) and (\ref{eq:dif_Gamma_full}), 
only the value of $f$ on the surface $J = J_{\rm cr}$ determines the comet injection rate. 
We characterize $f$ in this limit by examining the radial distribution of survivors 
with periastron distances $q_{\rm fin} \leq 1000 \AU$. 
Figure \ref{fig:PostKickPowerLaw} shows that, over a large range of $a_{\rm fin}$, 
the survivors follow power-law density profiles with $\alpha \approx 4$. 
This represents a slightly more concentrated profile than the initial state.

Overall, we find that the distribution of retained comets in our simulations 
can be well approximated as (cf.\ Eq.\ \ref{eq:fdist_fiducial})
\begin{equation} \label{eq:fdist_modified}
    f(L,J) =
    \begin{cases}
        C' N_{\rm c}' (L/L_{1})^{3-2\alpha}, & L_{1} \leq L \leq L_{\rm e}; \\
        C' N_{\rm c}' (L/L_{1})^{3-2\alpha}, & 0 \leq J \leq J_{\rm e}(L), L_{\rm e} \leq L \leq L_{2};  \\
        0, & \text{else;} 
    \end{cases}
\end{equation}
where
\begin{equation}
    J_{\rm e}(L) = L_{\rm e} \left( 2 - \frac{L_{\rm e}^{2}}{L^{2}} \right)^{1/2}, \hspace{0.5cm} L_{\rm e} = (G M q_{\rm e})^{1/2}
\end{equation}
The quantity $N_{\rm c}'$ is the total number of retained comets, i.e.\ $N_{\rm c}' = f_{\rm ret} N_{\rm c}$. 
This is the same distribution function used in Section \ref{s:OC-Dynamics} (Eq.\ \ref{eq:fdist_fiducial}), 
but it is truncated for $a \geq q_{\rm e}$ ($L \geq L_{\rm e}$) and $q \leq q_{\rm e}$ ($J \leq J_{\rm e}(L)$). 
The normalization factor $C'$ is a complicated function of $L_{1}$, $L_{2}$, and $\alpha$, 
but it reduces to Eqs.\ (\ref{eq:Cnorm}) in the limit $L_{0} \to L_{2}$. 
Numerically, $C'$ is only a factor of a few smaller than $C$ for a given $\alpha$.

\subsection{Revised bombardment rate} \label{s:MassLoss:Results:Flux-revisited}

Based on our characterization of the distribution function, 
the major assumptions made by \citetalias{HT1986} also hold 
for the retained comets in our simulations. 
We can therefore apply Eqs.\ (\ref{eq:dif_Gamma_empty}) and (\ref{eq:dif_Gamma_full}) as appropriate 
to calculate the comet accretion rate for the WD.

We evaluate the comet accretion rate after stellar mass loss
for the illustrative case $T = 1 \times 10^{5} \yr$ and $V_{\rm k} = 0.75 \, {\rm km \, s^{-1}}$. 
The remnant XOC contains $N'_{\rm c} \simeq 0.1 N_{\rm c}$ 
between $a_{1}' \simeq 2 \times 10^{3} \AU$ and $a'_{2} \simeq 8 \times 10^{4} \AU$ 
with $\alpha \simeq 4$. 
Using the \citetalias{HT1986} formalism with $f(L,J)$ given by Eq.\ (\ref{eq:fdist_modified}), 
we find 
\begin{equation}
    \dot{M}_{Z} \simeq 9 \times 10^{6} \, {\rm g \, s^{-1}} \left( \frac{f_{\rm acc} m_{\rm c}}{10^{17} \, {\rm g}} \right) \left( \frac{f_{\rm ret} N_{\rm c}}{10^{10}} \right).
\end{equation}
This is lower than the accretion rate calculated in Section \ref{s:OC-Dynamics:Gtide:Flux-at-WD} (Eq.\ \ref{eq:MdotZ_naive}). 
However, it still falls well within the range of measurable WD accretion rates. 
We conclude that the dynamical effects of late-stage stellar evolution 
cannot entirely explain the absence of cometary pollution. 

\section{Comet deflection by surviving planets} \label{s:Planets}

In Sections \ref{s:OC-Dynamics} and \ref{s:MassLoss},
we estimated the rate at which XOC comets are injected to within $q_{\rm cr} = r_{\rm tide}$ 
around a WD by the Galactic tide. 
However, the  presence of remnant planetary systems around WDs 
could potentially alter the trajectories of incoming comets, preventing their accretion by the WD. 
In the Solar System, LPCs have a high probability 
of being diverted by Jupiter or Saturn during each perihelion passage \citep[e.g.][]{Fernandez1981}. 
This `Jupiter--Saturn barrier' has long been thought to reduce the flux of LPCs near Earth 
and, by extension, the rate of LPC impacts on the Sun. 
In this section, we investigate the conditions under which surviving planets 
would prevent pollution of the star by XOC comets. 

\subsection{Orbit-averaged interactions} \label{s:Planets:Secular}

We first consider the case where the periastron distance $q = a(1-e)$ of an XOC comet 
is larger than the orbital semi-major axis of a perturbing planet ($a_{p}$). 
The planet--comet interaction may be averaged over the orbital motions of both bodies. 
The orbit of the comet therefore evolves due to both interior (planetary) and exterior (Galactic) secular forcing. 
This secular interaction gives rise to apsidal precession of the comet at a rate
\begin{equation}
    \left( \frac{\dif \chi}{\dif t} \right)_{p} = \frac{3}{8} \frac{m_{p} a_{p}^{2}}{M a^{2}} \frac{5 (\nhat \cdot \nhat_{p})^{2} - 1}{(1-e^{2})^{2}} \, \omega,
\end{equation}
where $m_{p}$ is the planet's mass 
and $\nhat$ and $\nhat_{p}$ are the comet's and planet's unit angular momentum vectors.
The Galactic tidal field also induces apsidal precession  (see Eq.\ \ref{eq:GT_EqsMot}). 
Competition between the planetary and Galactic tidal perturbations impose 
a maximum eccentricity $e_{\rm max}$ for the orbit of the comet \citep[e.g.][]{LML2015, PML2017} --
or, equivalent, a minimum periastron distance $q_{\rm min}$. 
This $e_{\rm max}$ can be estimated by equating the orders of magnitude 
of the apsidal precession rate due to each perturbation. 
This yields
\begin{equation}
    1 - e_{\rm max} \sim \left( \frac{m_{p} a_{p}^{2}}{\rho_{\rm g} a^{5}} \right)^{2/3}
\end{equation}
or, in terms of periastron distance,
\begin{align}
    q_{\rm min} &= a (1 - e_{\rm max}) \nonumber \\
    &\sim 20 \AU \left( \frac{m_{p}}{\MJ} \right)^{2/3} \left( \frac{a_{p}}{10 \AU} \right)^{4/3} \nonumber \\ 
    & \hspace{0.5cm} \times \left( \frac{\rho_{\rm g}}{0.1 \MSol \pc^{-3}} \right)^{-2/3} \left( \frac{a}{10^{4} \AU} \right)^{-7/3}. \label{eq:qmin_secular}
\end{align}
For a multi-planet system, the quantity $m_{p} a_{p}^{2}$ can be replaced with $K = \sum_{p} m_{p} a_{p}^{2}$, 
where the summation is taken over all planets. 

For a given $m_{p}$ and $a_{p}$, comets with sufficiently large $a$ have $q_{\rm min} \lesssim a_{p}$, 
i.e.\ they will still cross the orbit of the planet. 
Thus, from Eq.\ (\ref{eq:qmin_secular}), we see that planets with $m_{p} \gtrsim 1 \MJ$ and $a_{p} \gtrsim 10 \AU$
can repel incoming comets through orbit-averaged perturbations. 

\subsection{Direct scattering} \label{s:Planets:Scattering}

\begin{figure}
    \centering
    \includegraphics[width=\columnwidth]{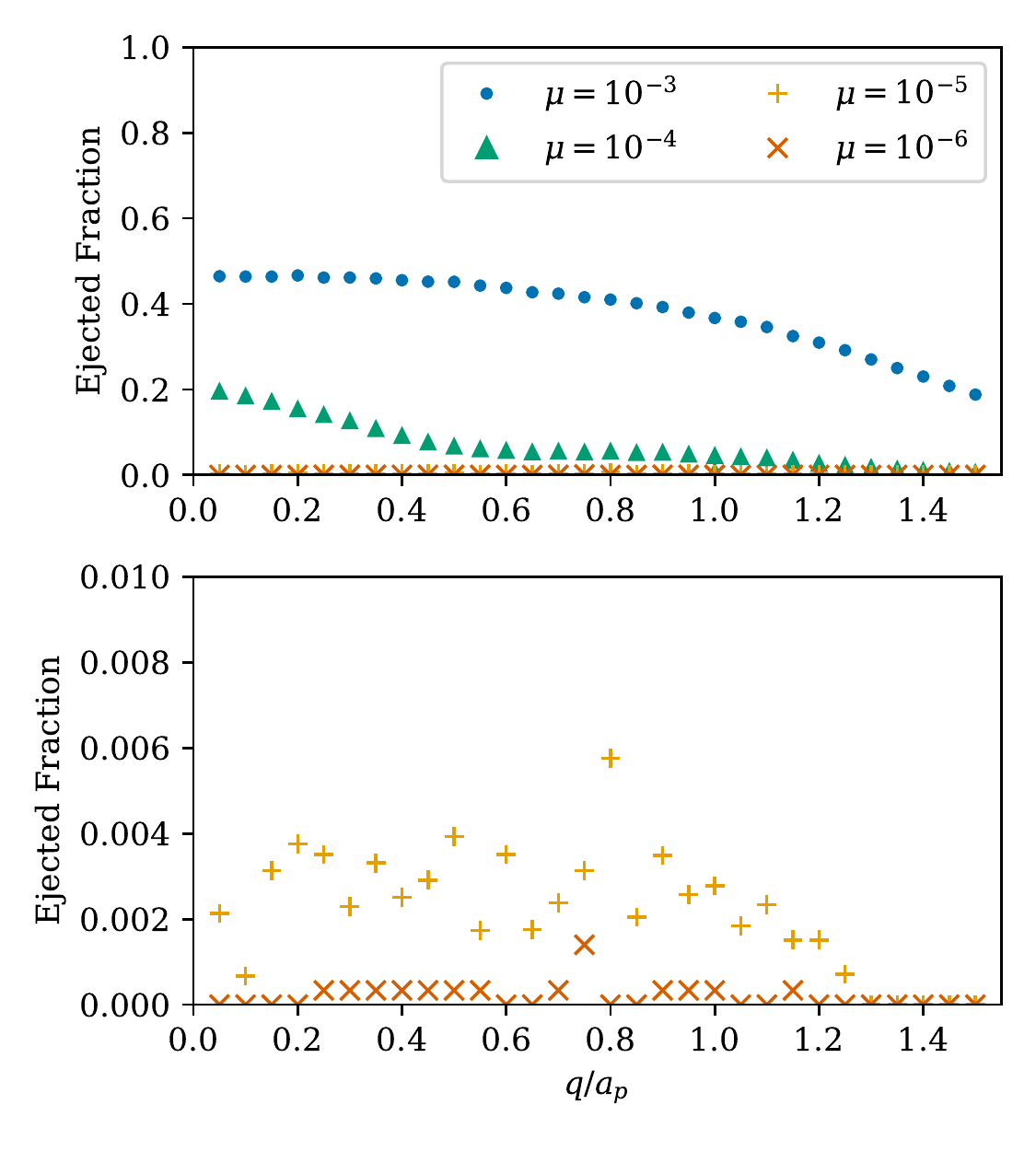}
    \caption{{\it Top:} Fraction of  comets 
    ejected from a single-planet system as a function 
    of $\mu = m_{p}/M$ and $q/a_{p}$. 
    The comets are initialized with $a = 2 \times 10^{4} \AU$. 
    {\it Bottom:} A vertical zoom on the data for $\mu = 10^{-5}, 10^{-6}$.}
    \label{fig:frac_pl_eject}
\end{figure}

A comet passing through a planetary system may be scattered after encountering a planet. 
The majority of comets do not experience a close encounter, 
defined as occurring when the minimum separation between the comet and planet 
is smaller than the Hill radius of the planet ($\sim a_{p} [m_{p}/M]^{1/3}$). 
However, the energy received by the comet from a distant encounter ($\sim G m_{p} / a_{p}$) 
can still be comparable to its orbital binding energy ($\sim G M / a$). 
The ratio of these quantities,
\begin{equation} \label{eq:def_Lambda}
    \Lambda  = \left( \frac{m_{p}}{\MJ}\,\bigg)\,\bigg( \frac{\MSol}{M} \right)\, \left( \frac{a}{10^{4} \AU}\,\bigg)\,\bigg( \frac{10 \AU}{a_{p}} \right),
\end{equation}
is reminiscent of the Safronov number for planetesimal scattering. 
For $\Lambda \ll 1$, the comet receives a negligible kick 
and can undergo many repeated passages 
until it collides with the planet or star. 
For $\Lambda \gtrsim 1$, the comet may be ejected into interstellar space or captured in a short-period orbit. 
Either way, incoming comets with $\Lambda \gtrsim 1$ may be efficiently removed from the OC 
and no longer have their $q$ reduced by the Galactic tide. 

To corroborate the general picture presented above, we conducted N-body simulations 
of the passage of a near-parabolic comet through a planetary system with {\sc rebound}. 
We considered a system containing a WD ($M = 0.6 \MSol$) 
and a single planet of mass $m_{p}$, semi-major axis $a_{p} = 10 \AU$, 
and eccentricity $e_{p} = 0.2$ (typical of long-period exoplanets). 
The comet was treated as a massless test particle, 
initialized on an incoming near-parabolic orbit 
with barycentric semi-major axis $a = 2 \times 10^{4} \AU$. 
We varied the comet's initial pericentre distance $q$, inclination $I$ (relative to the planetary orbit), 
apsidal angle $\chi$, and nodal angle $\Omega$, 
as well as the planet's initial mean longitude $\lambda_{p}$. 
We considered values of the planet-to-star mass ratio $\mu = m_{p}/M$ of $10^{-6}$, $10^{-5}$, $10^{-4}$, and $10^{-3}$. 
These correspond to $\Lambda$ values from $0.002$ to $2.0$. 

For each combination of $m_{p}$ and $q$, we conducted 45\,000 {\sc rebound} simulations, 
sampling $\cos{I}$, $\chi$, $\Omega$, and $\lambda_{p}$ from a uniform grid over their full range of possible values. 
We also sampled 30 uniformly spaced $q$ values on the interval $[0.005, 1.5] a_{p}$.
Each run lasted a time $2 T_{q}$, where $T_{q}$ is the time it would take 
for the comet to travel from its initial position to periastron 
along its initial orbit. 
We used the IAS15 integrator in {\sc rebound}. 

\begin{figure}
    \centering
    \includegraphics[width=\columnwidth]{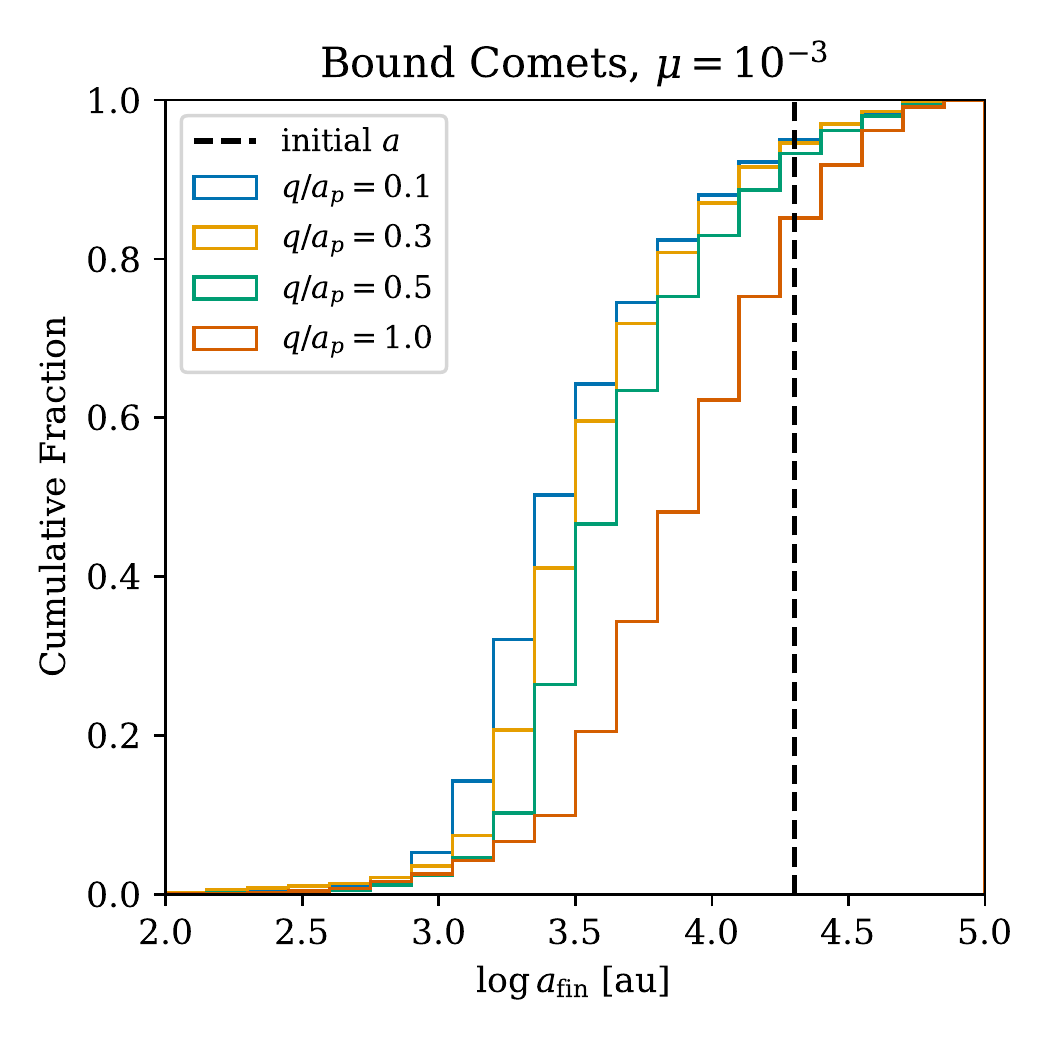}
    \caption{Cumulative distribution of the final semi-major axes of comets on bound orbits after a parabolic encounter with a WD--planet system with $a_{p} = 10 \AU$ and $\mu = 10^{-3}$. 
    The vertical dashed line indicates the comet's initial semi-major axis ($a = 2 \times 10^{4} \AU$).}
    \label{fig:dist_afin_1e-3}
\end{figure}

We determined the fate of each comet in post-processing. 
We assume that the comet is ejected if its final barycentric distance is at least $2 a_{p}$ 
and its final osculating orbit is either hyperbolic or elliptical with apoastron distance $Q > 0.8 \pc$. 
If neither condition was met, the comet was considered bound. 
In reality, the bound comets would either return to the OC or be captured on a shorter-period orbit. 
Either way, their ultimate fates are not determined in our simulations. 

Figure \ref{fig:frac_pl_eject} shows the fraction of incoming comets 
that are ejected as a function of initial $q$. 
These results are averaged over all other initial conditions 
(i.e.\ over the orientation of the comet's orbit). 
The ejected fraction generally increases with increasing $\mu$ and decreasing $q$. 
For $\mu = 10^{-6}$ and $10^{-5}$, only a very small fraction of comets are ejected, if any. 
For $\mu = 10^{-4}$, the fraction increases 
from $0.005$ at $q/a_{p} = 1.5$ to $0.19$ at $q/a_{p} = 0.005$.
For $\mu = 10^{-3}$, it increases from $0.18$ to $0.45$ over the same range. 

\begin{figure}
    \centering
    \includegraphics[width=\columnwidth]{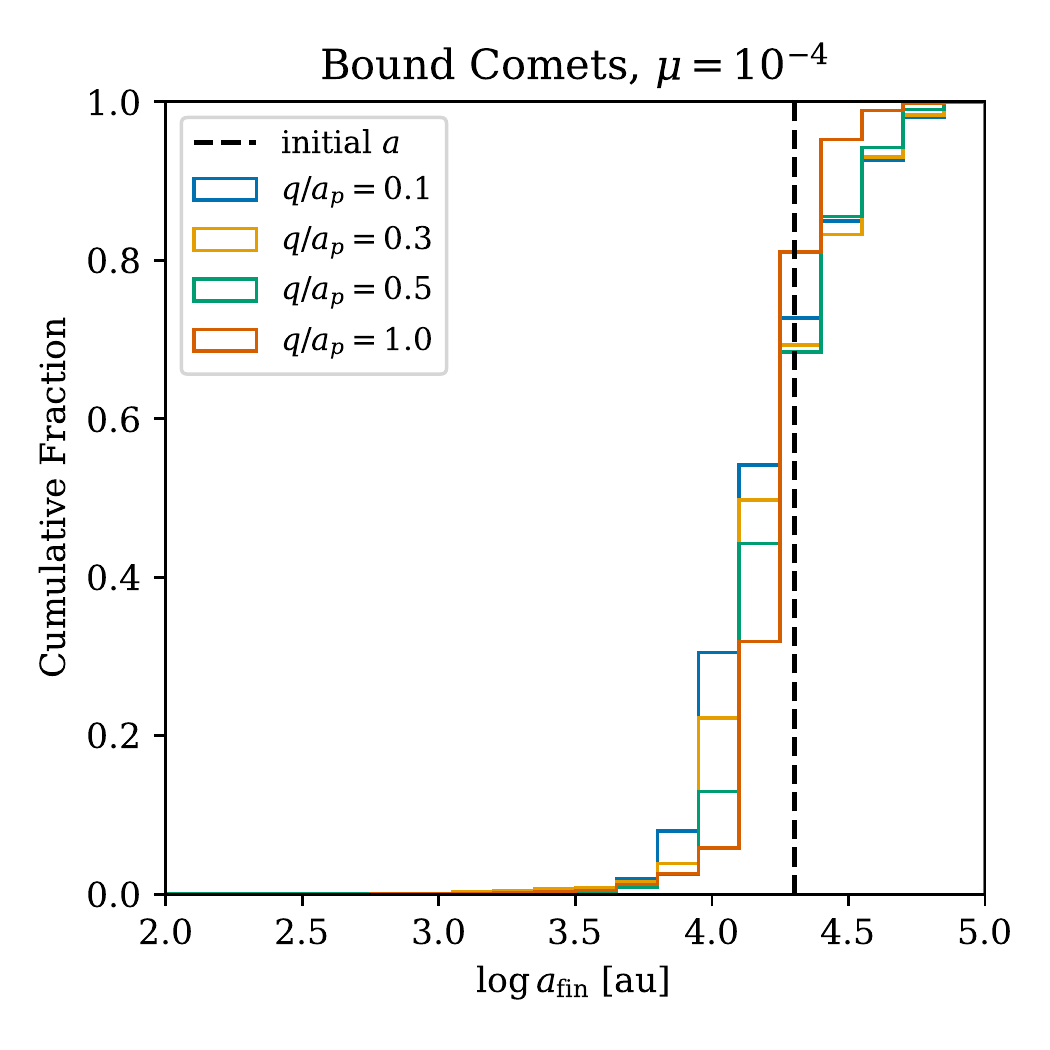}
    \caption{The same as Fig.\ \ref{fig:dist_afin_1e-3} with $\mu = 10^{-4}$.}
    \label{fig:dist_afin_1e-4}
\end{figure}

Figures \ref{fig:dist_afin_1e-3} and \ref{fig:dist_afin_1e-4} show cumulative distributions 
of the final semi-major axes of comets that remained bound after encountering the planet. 
This distribution is necessary to determine the fraction of bound comets 
that are continually pushed inward by the Galactic tide over successive periastron passages. 
We focus on the cases $\mu = 10^{-3}$ and $10^{-4}$, since the effect of the planet is negligible for smaller $\mu$. 

For $\mu = 10^{-3}$, 60 to 80 per cent of bound comets 
experience a reduction in semi-major axes by more than a factor of 2. 
A greater fraction of bound comets belong to this subset as $q$ decreases. 
Comets with $a_{\rm fin} \lesssim 3 \times 10^{3} \AU$ 
are decoupled from the Galactic tide (tidal torque $\propto a^{2}$) 
and no longer have their $q$ reduced between successive periastron passages. 
In the case of $\mu = 10^{-4}$, less than 20 per cent of bound comets
experienced a significant reduction of the semi-major axis. 
Moreover, almost none had $a_{\rm fin} \lesssim 3 \times 10^{3} \AU$. 
These results confirm qualitative expectations based on the value of $\Lambda$.

The statistical outcomes of near-parabolic encounters 
between a test particle and a star--planet system 
would be worthwhile to study in a future work. 
For the purposes of this work, we conclude that a planet with $\Lambda \gtrsim 1$
likely creates a dynamical barrier to LPCs with $q \lesssim a_{p}$. 
In the next section, we estimate the reduction of the  tidal disruption rate of LPCs from this effect. 

\subsection{Reduced comet injection rate}

OC comets interacting with Jupiter or Saturn have $\Lambda \gtrsim 1$. 
Consequently, it is generally thought that the Jovian planets reduce the overall flux of LPCs 
in the innermost part of the Solar System. 
However, to our knowledge, previous studies have not quantified 
the expected reduction of the LPC flux at very small $q$ due to scattering by planets. 
This is necessary to calculate the comet accretion rate of a planet-hosting WD. 
Incidentally, it also relates to the production of Sun-grazing comets from the OC
\citep[e.g.][]{Novski+2012, Fernandez+2021} 
and observed exocomets around planet-hosting main-sequence stars \citep[e.g.][]{Ferlet+1987}. 

Suppose that a comet has a probability $p_{\rm ej}$ of being ejected 
during each passage through the remnant planetary system of a WD.
The probability of being ejected after exactly $k$ passages is
\begin{equation}
    p(k) = p_{\rm ej} (1-p_{\rm ej})^{k-1},
\end{equation}
and the expected passage number on which ejection occurs is
\begin{equation}
    \langle k \rangle = \sum_{k=1}^{\infty} k p(k) = \frac{1}{p_{\rm ej}}\,.
\end{equation}
The numerical experiments presented in Section \ref{s:Planets:Scattering} demonstrated 
that $p_{\rm ej}$ approaches $0.5$ for planets with $\Lambda \gtrsim 1$. 
Therefore, in the presence of this dynamical barrier, 
the cumulative comet ejection fraction of LPCs with $q \lesssim a_{p}$ is $0.5$ after the first periastron passage,
$0.75$ after the second, and so on. 

For $r_{\rm tide} \leq q \lesssim a_{p}$, comets can be 
either injected by GTs 
to smaller $q$ (perhaps leading to tidal disruption) 
or ejected into interstellar space by the planet. 
The competition between these processes determines 
the comet accretion rate of the WD in the presence of the planet. 
A similar effect in dense star clusters with central black holes 
has been dubbed ``loss cone shielding'' by \citet{Teboul+2022}. 

\begin{figure}
    \centering
    \includegraphics[width=\columnwidth]{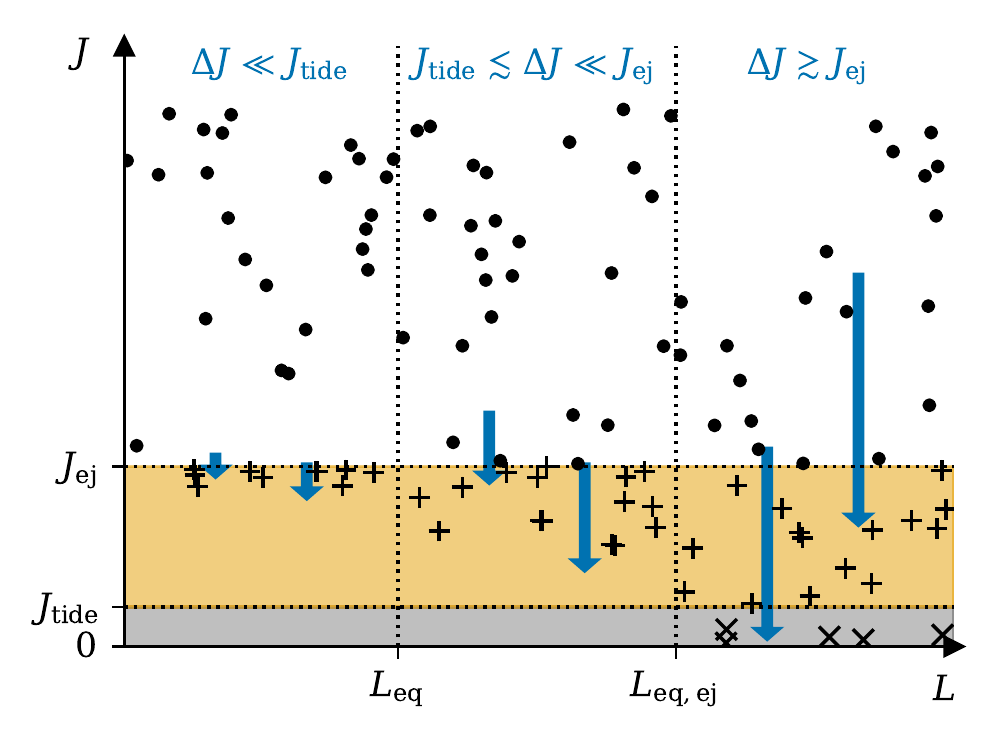}
    \caption{Schematic depiction of the modified loss cone theory in the presence of a planetary barrier. 
    The `tidal disruption' loss cone is shaded in grey, 
    whilst the `ejection loss cone' is shaded in light orange. 
    Comets with $L \lesssim L_{\rm eq,ej}$ are predominantly ejected by the planet (plus signs), 
    whilst those with $L \gtrsim L_{\rm eq,ej}$ are either ejected or tidally disrupted by the WD (crosses). 
    Note that the size of $J_{\rm tide}$ relative to $J_{\rm ej}$ is exaggerated for visual clarity.}
    \label{fig:losscone_2cones}
\end{figure}

We quantify the amount by which the accretion rate is reduced 
via a straightforward modification of the loss cone theory, 
illustrated schematically in Figure \ref{fig:losscone_2cones}.
We define an `ejection loss cone' with $J_{\rm cr} = (2 G M a_{p})^{1/2} \equiv J_{\rm ej}$; 
in general, we have $J_{\rm ej} \gg J_{\rm tide} = (2 G M r_{\rm tide})^{1/2}$. 
The two loss cones overlap, with the ejection loss cone 
covering a much larger region of phase space. 
For a comet with relatively small $a$ ($L$), such that the change of its angular momentum per orbit $\Delta J$ 
is smaller than $J_{\rm ej}$, the ejection loss cone is empty; 
for a comet with larger $a$ such that $\Delta J \gtrsim J_{\rm ej}$, this loss cone is filled. 
The transition semi-major axis is
\begin{equation} \label{eq:def_aeq_ejection}
    a_{\rm eq,ej} \simeq 0.38 \left( \frac{M^{2} a_{p}}{\rho_{\rm g}^{2}} \right)^{1/7},
\end{equation}
which exceeds $a_{\rm eq}$ (see Eq.\ \ref{eq:afill_tides}) by a factor $(a_{p} / r_{\rm tide})^{1/7}$; 
we also define the corresponding value $L_{\rm eq,ej} \equiv (G M a_{\rm eq,ej})^{1/2}$. 
Because the tidal-disruption loss cone is nested deep inside the ejection loss cone, 
comets with $L \lesssim L_{\rm eq,ej}$ are mostly ejected before they reach the tidal radius ($J = J_{\rm tide}$). 
On the other hand, those with $L \gtrsim L_{\rm eq,ej}$ can reach $J \leq J_{\rm tide}$ in a single orbit, avoiding ejection. 
The accretion rate onto the WD from comets in the filled ejection loss cone 
is therefore given by the steady-state number of comets with $J \lesssim J_{\rm tide}$ (at a given $L$) 
divided by their orbital period $P(L)$ -- 
the same as Eq.\ (\ref{eq:dif_Gamma_full}) with $J_{\rm cr} = J_{\rm tide}$. 
Thus, in the presence of a strong planetary barrier, the total comet bombardment rate on the WD is
\begin{equation}
    \Gamma_{\rm tot}^{\rm (pl)} \simeq \int_{L_{\rm eq,ej}}^{L_{2}} \Gamma_{\rm f}(L) \, \dif L \sim L_{\rm eq,ej} \Gamma_{\rm f}(L_{\rm eq,ej}).
\end{equation}
This can be compared to the bombardment rate without a planetary barrier, $\Gamma_{\rm tot}^{\rm (no \ pl)} \sim L_{\rm eq} \Gamma_{\rm f}(L_{\rm eq})$ (see Eq.\ \ref{eq:Gamma_tot}). 
Using the relations $\Gamma_{\rm f}(L) \propto f(L,J_{\rm tide}) / L^{3}$ (Eq.\ \ref{eq:dif_Gamma_full}) and $f \propto L^{3-2\alpha}$ (Eq.\ \ref{eq:fdist_modified}), 
we find
\begin{equation}
    \frac{\Gamma_{\rm tot}^{\rm (pl)}}{\Gamma_{\rm tot}^{\rm (no \ pl)}} \sim \left( \frac{L_{\rm eq,ej}}{L_{\rm eq}} \right)^{1 - 2\alpha} = \left( \frac{r_{\rm tide}}{a_{p}} \right)^{(2\alpha-1)/14}.
\end{equation}
For typical values $2 \leq \alpha \leq 4$, this ratio is much less than unity unless $a_{p}$ is close to $r_{\rm tide}$. 
For example, taking $\alpha = 7/2$, we have
\begin{equation}
    \frac{\Gamma_{\rm tot}^{\rm (pl)}}{\Gamma_{\rm tot}^{\rm (no \ pl)}} \sim \left( \frac{r_{\rm tide}}{a_{p}} \right)^{3/7} \approx 0.05 \left( \frac{r_{\rm tide}}{0.01 \AU} \right)^{3/7} \left( \frac{a_{p}}{10 \AU} \right)^{-3/7} .
\end{equation}
We conclude that planets with $\Lambda \gtrsim 1$ and $a_{p} \gg r_{\rm tide}$ 
are capable of significantly reducing the comet accretion rate of a WD. 
Notice that the planetary barrier reduces the accretion rate more severely 
when the XOC is more centrally concentrated (greater $\alpha$).

\section{Discussion} \label{s:Discussion}

We find that both late-stage stellar evolution and the presence of a surviving planetary system 
can reduce the rate at which XOC comets bombard a central WD. 
The primary cause of the reduction is the ejection of comets into interstellar space. 
This can occur either during stellar mass loss or after a scattering event from one of the planets. 
The simplified model XOC described in Section \ref{s:OC-Dynamics}
produces a metal accretion rate given by:
\begin{equation*}\label{eq:mdot}
    \dot{M}_{Z} \sim 10^{8} \, {\rm g \, s^{-1}} f_{\rm acc} \left( \frac{m_{\rm c}}{10^{17} \, {\rm g}} \right) \left( \frac{N_{\rm c}}{10^{11}} \right),
\end{equation*}
where $N_{\rm c}$ refers to the number of comets in the cloud with $a > 10^{4} \AU$ 
and $m_{\rm c}$ is the typical mass of a comet. 
The results of the previous two sections can be incorporated into this estimate
by multiplying the right-hand side of Eq.\ (\ref{eq:mdot}) by two additional factors, $f_{\rm ret}$ and $f_{\rm bar}$. 
These two factors account for the retention of comets through stellar evolution 
and the planetary barrier effect. 
We found that each of these factors is typically $\approx 0.01$--$0.1$, 
depending on the magnitude of WD kicks ($f_{\rm ret}$) 
and the orbital radii of the surviving planets ($f_{\rm bar}$). 
Neither process acting alone is sufficient to reduce the comet accretion rate
below the observational detection threshold of $\sim 10^{5} \, {\rm g \, s^{-1}}$. 
However, both processes operating in parellel could reduce the accretion rate below the observational constraints. 
An inefficient accretion process ($f_{\rm acc} \lesssim 1$) would further reduce $\dot{M}_{Z}$.

The factors $f_{\rm ret}$ and $f_{\rm bar}$ are expected to vary between systems. 
A subset of polluted WDs could therefore be accreting 
marginally detectable amounts of cometary debris ($\sim 10^{5}$--$10^{6} \, {\rm g \, s^{-1}}$). 
Given that only a few dozen polluted WDs have been characterized in detail, 
these may have gone unnoticed to date. 
A confounding factor is that comet-hosting WDs may also possess populations 
of intrinsically `rocky' bodies, such as exo-asteroids. 
Many authors have argued that these are the main parent-body population 
for the observed pollution in many systems 
\citep{Jura2006, Jura2008, DWS2012, Mustill+2018, Smallwood+2018, Smallwood+2021, OTL2022, Trierweiler+2022}. 
If exo-asteroids are accreted by the WD at a much greater rate than icy bodies on average, 
then the contribution from exocomets to the atmospheric chemical abundances may be obscured.

\subsection{Comparison with related works} \label{s:Discussion:Compare}

\citet{AFS1986} and \citet{PA1998} respectively carried out early studies of 
the dynamical effects of symmetric and asymmetric mass loss on an XOC.
They used a parametrized prescription for the mass loss and natal kick similar to ours.  
The fraction of comets that remain bound to the WD in their simulations 
agrees with our result in order of magnitude.

\citet{Veras+2014} conducted N-body simulations of the evolution of an XOC 
during late-stage stellar evolution for host stars of several different masses. 
Their simulations included the Galactic tidal field and random stellar flybys 
but did not include a possible surviving planetary system or a natal kick on the host. 
Despite this, \citeauthor{Veras+2014} found a lower comet accretion rate 
compared to our analytical estimate: 
specifically, they found that a fraction $\sim 10^{-5}$ of comets 
were accreted by the WD over $10 \Gyr$. 
When assuming a $\sim 1 \ME$ cloud, this corresponds to an average accretion rate 
of $\dot{M}_{Z} \sim 10^{5} \, {\rm g \, s^{-1}}$; this rate is comparable to the observational detection limit 
and far below the value $\dot{M}_{Z} \sim 10^{8} \, {\rm g \, s^{-1}}$ 
we estimated analytically. 
The difference most likely arises because \citeauthor{Veras+2014} 
considered OC comets initially distributed between $10^{4}$ and $10^{5} \AU$ on the MS, 
rather than from $10^{3}$ to $10^{5} \AU$ as in our case.
A larger fraction of the comets would have been lost during stellar mass loss 
in their simulations than in ours, reducing the subsequent accretion rate of the WD. 

\citet{SML2015} studied the direct injection of cometary material to $\sim 10 \AU$ distances
as a result of the natal kick of the WD. 
They calculated the change of comet orbits using an impulse approximation throughout the cloud. 
Specifically, they assumed that the kick is imparted in the final few $10^{4} \yr$ of stellar evolution, 
after the vast majority of mass loss has occurred through an isotropic wind. 
This may be more realistic than our assumption that the mass loss and kick occur simultaneously, 
and the distribution function of surviving comets would likely be different. 
However, \citeauthor{SML2015} did not discuss the long-term pollution of the central WD in their study.

Finally, \citet{CH2017} calculated the orbital evolution of XOC comets during isotropic stellar mass loss, 
using a realistic time-dependent mass loss rate from {\sc mesa}. 
They did so by constructing an approximate interpolation scheme 
between the adiabatic and impulsive mass-loss regimes. 
This scheme, or alternatively that of \citet{Veras+2011}, 
could be adapted to include a time-dependent kick acceleration on the host star 
in order to predict the distribution function of surviving comets more accurately.
This is outside of the scope of this paper but would be useful to investigate in future work.

\subsection{Comets and habitability of WD planets} \label{s:Discussion:Habitability}

The discovery of surviving short-period planets around WDs \citep[e.g.][]{Vanderburg+2020}
has prompted speculation and investigation regarding their habitability. 
Particular emphasis has been placed on hypothetical Earth-sized rocky planets (\citealt{Kaltenegger+2020} and references therein). 
Comet impacts are integral to the question of planetary habitability in the Solar System and elsewhere. 
Comets can deliver the chemical ingredients of terrestrial life 
-- water and other volatile compounds --
to worlds formed inside the condensation lines of these species in the birth nebula 
\citep[e.g.][]{OwenBarNun1995, Dauphas+2000, Dauphas2003}. 
Impact events by either comets or asteroids are a potential cause of 
mass extinctions or climatic shifts on our own planet \citep[e.g.][]{Alvarez+1980, Kent+2003}.

Here we focus on the ability of comets to deliver volatile substances. 
The traditional habitable zone around an evolved, cool WD 
occurs at $a \approx 0.01 \AU$ \citep{Agol2011}.
Either dynamical migration \citep{VG2015, MP2020, OLL2021, Stephan+2021} or 
second-generation formation \citep{BS2015, vL+2018} is required for rocky planets to exist there. 
Such planets are prone to volatile depletion through intense tidal heating \citep{BH2013} 
or ultraviolet irradiation \citep{Lin+2022}. 
Delivery of XOC comets to short-period orbits and subsequent collisions with planets could, in principle, replenish the volatiles. 

The habitable zone and the Roche limit occur at roughly the same distances 
around an old WD ($\simeq 0.01 \AU$), 
notwithstanding extra heating processes such as tidal friction \citep{Becker+2023}. 
Therefore, the rate at which a short-period planet captures cometary material 
is at most of the order of the accretion rate of the host star. 
The fraction of material captured by a planet via direct impacts  depends on 
its accretion cross-section and the comet's orbital geometry \citep{Torres+2021, Seligman+2022}.
A planet near the Roche limit may be able to capture additional material 
from a circumstellar debris disc \citep{vL+2018}.

Based on the non-detection of comet-like debris in polluted WD atmospheres, 
we place a characteristic upper bound of $\dot{M}_{Z} \approx 10^{5} \, {\rm g \, s^{-1}}$
for the time-averaged accretion rate of cometary material 
onto hypothetical short-period planets. 
This constrains the total amount of volatiles that these planets 
can acquire from comet impacts over their lifetimes. 
For example, if the mass fraction of water ice in a typical comet is $X_{\rm H2O}$, 
then the planet can accrete water from comets at a maximum rate
\begin{align}
    \dot{M}({\rm H_{2}O}) \approx 10^{21} \, {\rm g} \Gyr^{-1} \left( \frac{\dot{M}_{Z}}{10^{5} \, {\rm g \, s^{-1}}} \right) \left( \frac{X_{\rm H2O}}{0.3} \right).
\end{align}
Based on theoretical WD cooling models, \citet{Kozakis+2018} found 
that a planet orbiting a $0.6 \MSol$ WD at $a = 0.01 \AU$ can remain habitable for $[4,9] \Gyr$. 
For $\dot{M}_{Z} = 10^{5} \, {\rm g \, s^{-1}}$ and $X_{\rm H2O} = 0.3$, 
this implies up to $[4,9] \times 10^{21} \, {\rm g}$ of cometary water 
may be accreted by such a planet during its habitable lifetime. 
This is much less than the mass of Earth's oceans, 
but it is comparable to that of Mars' ice caps \citep{Christensen2006}. 
Similarly, for a bulk nitrogen mass fraction $X_{\rm N} = 0.015$ in a Halley analogue \citep{Jessberger+1988}, 
some $[2,5] \times 10^{20} \, {\rm g}$ of N 
could be delivered in the same timeframe. 
This would amount to $[4,10]$ per cent of the total mass of Earth's atmosphere.
In short, accretion from comets at the level of current observational constraints 
could provide a rocky planet in the habitable zone of a WD 
with a significant global budget of volatile compounds. 

\subsection{Caveats} \label{s:Discussion:Caveats}

\subsubsection{Axisymmetric approximation of the GT} \label{s:Discussion:Caveats:QuadGT}

The comet injection rate calculated by \citetalias{HT1986} only accounted for the vertical component of the GT. 
The radial and tangential components are smaller by an order of magnitude. 
The dynamics of a comet under the vertical GT is an instance 
of the evolution of a binary in an axisymmetric external potential \citep{HR2019a}. 
The inclusion of weaker non-axisymmetric components tends 
to excite extreme orbital eccentricities \citep[e.g.][]{Naoz+2013, Pejcha+2013}.
Indeed, \citet{Fouchard+2006} found that the radial tidal force slightly increases 
the comet delivery rate in the Solar System at $q \sim$ few au. 
Similar effects may enhance the accretion rate of a WD.
It would be worthwhile to generalize the calculation 
of the injection rate in \citetalias{HT1986} to the non-axisymmetric case in future work.

\subsubsection{Isotropy and scale of XOCs} \label{s:Discussion:Caveats:Isotropy}

We have assumed that the XOC has a spherically symmetric structure throughout this work. 
Here we briefly qualify this assumption  to order-of-magnitude accuracy.

The outer OC surrounding the Solar System is likely mildly anisotropic  
due to the anisotropy of the Galactic tidal field \citepalias[e.g.][]{HT1986}. 
The inner cloud is likely highly anisotropic, due to its
presumed dynamical relationship with the giant planets \citep[e.g.][]{Duncan+1987}. 
As we demonstrated in Section \ref{s:MassLoss}, comets in the inner OC 
are preferentially retained through late stellar evolution. 
Therefore the surviving XOC of a young WD may inherit 
the same degree of anisotropy, 
modulo the effects of asymmetric mass loss. 
It is feasible to incorporate this anisotropy in our model
by truncating the fiducial spherical distribution function for a range of orientations. 
It would then be straightforward to calculate the comet injection rate 
from such a cloud following similar steps to \citetalias{HT1986}. 

On the other hand, the degree of anisotropy in a cloud would be 
reduced over time following post-MS orbital expansion. 
This is because the expansion of cometary orbits 
makes them more susceptible to diffusion driven by weak stellar flybys. 
Based on theoretical models for the Solar System, 
a surviving XOC around a WD would become dynamically relaxed around a cooling age of $\sim 1 \Gyr$. 

Another underlying assumption of our discussion is that XOCs 
have the same characteristic size as the solar OC. 
However, the size of an XOC is a function of the architecture and initial conditions of the inner planetary system 
and the density of the galactic environment in which it formed \citep{Fernandez1997}. 
If the progenitors of single WDs form in denser environments  
or have more compact or less massive planetary systems, 
then their XOCs would be smaller and less susceptible to ejection during stellar mass loss. 
On the other hand, their comets would be less prone to injection by Galactic tides 
when the host star resides within a less dense environment.

\subsubsection{Composition of cometary nuclei} \label{s:Discussion:Caveats:Composition}

We have assumed throughout this work that cometary debris in a WD atmosphere 
would have bulk elemental abundances similar to Halley's comet
\citep[cf.][]{Jura2006, Xu+2017}. 
However, cometary nuclei within the Solar System display diverse compositions \citep[e.g.][]{AHearn+2012, Cochran+2015, BMB2017}. 
Notably, some have sub-solar C abundances on par with 
those measured in polluted WDs \citep{Seligman+2022}. 
It is plausible that some WDs are polluted by relatively volatile-poor comets.

The issue of composition is also related to the question of XOC scale 
raised in Section \ref{s:Discussion:Caveats:Isotropy}. 
A comet in a relatively compact XOC ($\lesssim 10^{3} \AU$) 
would be subjected to intense heating during the host's post-MS evolution, 
possibly leading to loss of volatiles by sublimation. 
In that case, the material that subsequently pollutes the WD may be dominated by refractory elements, 
perhaps rendering it indistinguishable from other rocky debris \citep{Zhang+2021}. 
G.\ Levine et al.\ (submitted) have addressed this by modelling the thermal evolution 
of OC objects around post-MS stars. 
They find that objects in the inner OC up to $\sim 1 \, {\rm km}$ in size
may become depleted in hypervolatiles 
(compounds with low sublimation temperatures, such as H$_{2}$, CO, and N$_{2}$) 
due to post-MS thermal processing (G.\ Levine, private communication). 
However, hypervolatiles constitute only a moderate fraction of the total mass in typical cometary nuclei;
compounds with relatively high sublimation temperatures (e.g.\ H$_{2}$O, CO$_{2}$, NH$_{3}$) 
are often present alongside hypervolatiles in similar amounts
(e.g.\ \citealt{Biver+2022} and references therein). 
Observations of polluted WDs reveal only the bulk elemental abundances of parent bodies, 
rather than the abundances of individual compounds. 
Thus, it is plausible that the chemical signature of an XOC object in a WD atmosphere 
would approximately reflect its primordial bulk composition, 
despite post-MS thermal processing.

\section{Conclusion} \label{s:Conclusion}

We have studied dynamical processes operating on extrasolar Oort clouds (XOCs)
in relation to the rate at which long-period comets are accreted by WDs. 
Our investigation has been informed by the observations that polluted WD atmospheres 
are typically depleted in volatile heavy elements. 
Our main conclusions are as follows:

\begin{enumerate}
    \item[(i)] Adapting the loss cone theory of comet injeciton by the Galactic tide for the Oort Cloud \citepalias{HT1986} 
    we show that a WD would accrete cometary debris from a Solar-System-like XOC 
    at a rate comparable to the typical total accretion rates observed in polluted systems. 
    Assuming the composition of Halley's comet is representative of XOC objects, 
    volatile elements would be readily observable in the WD's atmosphere in this scenario. 
    This conclusion is insensitive to the radial distribution of comets, 
    provided that the inner boundary of the cloud lies between $10^{3}$ and $10^{4} \AU$. 
    \item[(ii)] Post-MS stellar mass loss, including a possible natal kick imparted to a WD, 
    causes a majority of XOC comets to be ejected as free-floating bodies. 
    The fraction of comets retained by a newborn WD is sensitive to the kick speed 
    but insensitive to the time-scale of stellar mass loss (within a realistic range). 
    Based on current constraints on the typical WD kick, 
    single WDs retain between $1$ and $30$ per cent of XOC comets. 
    The orbital distribution function of the survivors is modified to a moderate degree.
    \item[(iii)] The presence of a surviving planetary system around a WD may reduce the rate at which comets fall to the Roche limit around a WD.
    Orbit-averaged perturbations from the planets on the comet's orbit
    can prevent comets interior to $\sim 10^{4} \AU$ 
    from reaching small periastron distances under the influence of the Galactic tide (see Section \ref{s:Planets:Secular}). 
    More importantly, direct scattering can cause up to $\sim 50$ per cent of comets to be ejected on hyperbolic orbits, 
    provided that the planet's mass $m_{p}$ and semi-major axis $a_{p}$ satisfy (see Eq.\ \ref{eq:def_Lambda})
    \begin{equation*}
        \Lambda = \frac{m_{p}}{M} \frac{a}{a_{p}} \gtrsim 1,
    \end{equation*}
    where $M$ is the WD's mass and $a$ is the comet's semi-major axis. 
    Using a modified loss cone theory, we show that the presence of a ``planetary barrier''
    can significantly reduce the fraction of comets that reach the Roche limit (see Section \ref{s:Planets:Scattering}).
\end{enumerate}

We find that stellar mass loss alone cannot sufficiently reduce the comet accretion rate 
to prevent observational detection of accreted volatiles. 
Therefore, to explain the dearth of detected volatiles in WDs, 
we suggest that a large fraction of polluted single WDs possess surviving planets with $\Lambda \gtrsim 1$ 
(with respect to a `standard' XOC size $\sim 10^{4} \AU$). 
Planets that avoided engulfment during post-MS evolution 
would orbit at distances $\gtrsim 5 \AU$ \citep[e.g.][]{MV2012}. 
In order to block incoming comets effectively at this distance, 
the mass ratio of the planet relative to the WD must be $\gtrsim 10^{-3}$. 
This suggests that the surviving planets have typical masses $\gtrsim 0.6 \MJ (M / 0.6 \MSol)$. 
This inference is consistent with (i) observationally inferred giant-planet occurrence rates around A- and F-type MS stars 
\citep[e.g.][]{Johnson+2010, Reffert+2015, Jones+2016, Ghezzi+2018} 
and (ii) indirect chemical evidence for the presence 
of evaporating giant planets around a large fraction of young WDs \citep{Schreiber+2019}.
Our prediction is testable in the near term through planet searches 
targeting WDs using direct imaging \citep{BZK2021,Mullally+2021_jwst}, 
astrometry \citep{SBM2022}, and microlensing \citep{Blackman+2021}.

The absence of volatile-enriched pollution among WDs has ramifications 
on the habitability and surface conditions of (hypothetical) short-period rocky planets orbiting these stars. 
Assuming that the accretion rate of cometary debris by such a planet 
is equal to the observational detection limit for the accretion rate of comets in the WD ($\sim 10^{5} \, {\rm g \, s^{-1}}$), 
global water and nitrogen budgets similar to present-day Mars and Earth, respectively, 
can both be achieved on Gyr time-scales. 

Near the completion of this work, we learnt of a similar study 
being carried out by D.\ Pham and H.\ Rein. 
They are developing numerical techniques to study 
the long-term orbital evolution of a large ensemble of XOC comets 
under the same dynamical effects we have considered (D.\ Pham \& H.\ Rein, private communication). 
Their approach is complementary to ours and will help to validate our analytical results.

\section*{Acknowledgements}

We thank Juliette Becker, Boris G\"{a}nsicke, Chris Hamilton, Garrett Levine, Siyi Xu, and Ben Zuckerman for helpful discussions, 
as well as the organizers and participants of the KITP program 
{\it White Dwarfs as Probes of the Evolution of Planets, Stars, 
the Milky Way and the Expanding Universe} in Fall 2022. 

This work has been supported in part by National Science Foundation Grants Nos.\ PHY-1748958 and AST-2107796, 
as well as the Heising-Simons Foundation and the Simons Foundation (216179, LB). 
CEO gratefully acknowledges a Space Grant Graduate Research Fellowship from the New York Space Grant Consortium. 
DZS acknowledges financial support from NSF Grant No.\ AST-2107796, 
NASA Grant No.\ 80NSSC19K0444, and NASA Contract NNX17AL71A. 

This work has made use of NASA's Astrophysics Data System and of the software libraries {\sc matplotlib} \citep{Hunter2007}, {\sc numpy} \citep{Harris+2020}, and {\sc scipy} \citep{Virtanen+2020}.

\section*{Data availability}

The data underlying this article will be shared on reasonable request to the corresponding author.

\bibliography{refs}
\bibliographystyle{mnras.bst}

\label{lastpage}

\end{document}